\documentclass[11pt,letterpaper]{article}
\usepackage{amsmath,amsthm,amsfonts,color,hyperref,anysize,amssymb,color,graphicx,epstopdf,mathtools,mathrsfs,bm}
\usepackage[section]{placeins}
\usepackage[usenames,dvipsnames]{xcolor}
\usepackage[normalem]{ulem}
\usepackage[ruled,vlined,linesnumbered]{algorithm2e}
\usepackage{algorithmicx}
\usepackage[margin=1.0in]{geometry}
\usepackage{multirow}
\usepackage{multicol}
\usepackage{subfiles}
\usepackage[shortlabels]{enumitem}
\usepackage{authblk}

\newcommand\numberthis{\addtocounter{equation}{1}\tag{\theequation}}
\newcommand{\hide}[1]{}
\newtheorem{theorem}{Theorem}[section]

\newtheorem{claim}[theorem]{Claim}
\newtheorem{lemma}[theorem]{Lemma}

\newtheorem{defi}[theorem]{Definition}
\newenvironment{claimproof}[1][\proofname]
	{  
        \proof[Proof]
        
	}
	{      
		\endproof
	}

\newif\ifnotes\notestrue

\ifnotes

\else

\fi

\newcommand{\Ai}{A_i}
\newcommand{\Bi}{B_i}

\newcommand{\bi}{b_i}
\newcommand{\Vm}{V^{\max}}
\newcommand{\Vmm}{\tilde V}
\newcommand{\Cm}{C^{\min}}
\newcommand{\bu}{\bar u}
\newcommand{\Hd}{H}

\newcommand{\R}{\mathbb{R}}
\newcommand{\budgeti}{w_i}
\newcommand{\edw}{{e}}
\newcommand{\edwij}{e_{ij}}
\newcommand{\edwi}{e_{i}}

\newcommand{\ti}{t_i}
\newcommand{\qi}{q_i}
\newcommand{\si}{s_i}

\newcommand{\pj}{p_j}

\newcommand{\C}{C}

\newcommand{\aijl}{a_{ij}^l}
\newcommand{\aijlp}{a_{ij'}^l}
\newcommand{\bil}{b_i^l}
\newcommand{\amax}{1}

\newcommand{\aij}{a_{ij}}

\newcommand{\ci}{\gamma_i}

\newcommand{\nameone}{\text{regular}}
\newcommand{\nametwo}{\text{robust}}
\usepackage{cleveref}

\title{Approximating Equilibrium under Constrained Piecewise Linear Concave Utilities with Applications to Matching Markets\thanks{The first author is supported by NSF Grant CCF--1942321 (CAREER). The second and third authors received funding from the European Research Council (ERC) under the European Union’s Horizon 2020 research and innovation programme (grant agreement no. ScaleOpt--757481).}}
\author[1]{Jugal Garg}
\author[2]{Yixin Tao}
\author[2]{L{\'{a}}szl{\'{o}} A. V{\'{e}}gh}
\affil[1]{University of Illinois at Urbana-Champaign\\  \texttt{jugal@illinois.edu}}
\affil[2]{London School of Economics and Political Science\\
\texttt{\{y.tao16,l.vegh\}@lse.ac.uk}}
\date{}

\begin{document}

\maketitle
\thispagestyle{empty}
\begin{abstract}

We study the equilibrium computation problem in the Fisher market model with constrained piecewise linear concave (PLC) utilities. 
This general class captures many well-studied special cases, including markets with PLC utilities, markets with satiation, and matching markets. For the special case of PLC utilities, although the problem is PPAD-hard, Devanur and Kannan (FOCS 2008) gave a polynomial-time algorithm when the number of items is constant. Our main result is a fixed parameter approximation scheme for computing an approximate equilibrium, where the parameters are the number of agents and the approximation accuracy. This provides an answer to an open question by Devanur and Kannan for PLC utilities, and gives a  simpler and faster algorithm for matching markets as the one by Alaei, Jalaly and Tardos (EC 2017).

 The main technical idea is to work with the stronger concept of thrifty equilibria, and approximating the input utility functions by `robust' utilities that have favorable marginal properties. With some restrictions, the results also extend to the Arrow--Debreu exchange market model.
\end{abstract}

\newpage
\setcounter{page}{1}

\section{Introduction}\label{sec:intro}

Market equilibrium is one of the most fundamental solution concepts in economics, where prices and allocations are such that demand meets supply when each agent gets her most preferred and affordable bundle of items. Due to the remarkable fairness and efficiency guarantees of equilibrium allocation, it is also one of the preferred solutions for fair division problems even though there may be no money involved in the latter case. A prominent example is competitive equilibrium with equal incomes (CEEI)~\cite{Varian74}, where a market is created by giving one dollar of virtual money to every agent.  

In this paper, we focus on markets with divisible items. Extensive work in theoretical computer science over the last two decades has led to a deep understanding of the computational complexity of equilibria for the classical models of Fisher and exchange markets, introduced by Fisher~\cite{BrainardS00} and Walras~\cite{Walras74} respectively in the late nineteenth century. In a Fisher market, agents have fixed budgets to spend on items according to their preferences given by utility functions over bundles of items. CEEI is a special case of this model, where each agent has a budget of one dollar. In the exchange (also known as Arrow--Debreu) market model, the items are brought to the market by the agents, who can spend their revenue from selling their initial endowments. 

Prevalent assumptions on the utility functions in the literature  are {\em (a)} \emph{monotonicity}, i.e., getting a bundle containing more of each item may not decrease the utility, and {\em (b)} \emph{local non-satiation}, i.e., for every bundle of items, an arbitrary neighborhood contains a bundle with strictly higher utility. A prominent example where these assumptions do not hold is the \emph{one-sided matching market problem}, where each agent needs to be assigned \emph{exactly} one unit of fractional items in total. Hylland and Zeckhauser~\cite{HyllandZ79} introduced an elegant mechanism based on CEEI for the one-sided matching markets. However, more general allocation constraints remain largely unexplored.

In this paper, we consider the equilibrium computation problem when agents have \emph{constrained piecewise linear concave (PLC)} utility functions, defined as follows.

\begin{defi}\label{def:cplc}
The utility function $u_i:\R_+^m\to \R\cup\{-\infty\}$ of agent $i$ is \emph{constrained PLC} if it is given in the following form. For some $p_i,r_i\in\mathbb{N}$, let $A_i\in\R^{p_i\times m}$, $B_i\in\R^{p_i\times r_i}$, $q_i\in\R^m$, $s_i\in \R^{r_i}$,
\[
u_i(x_i)=\begin{cases}
\max_{t_i} \qi^\top x_i + \si^\top \ti  \text{ s.t. } \Ai x_i + \Bi \ti\leq \bi &\text{ if }\exists t_i:\,  \Ai x_i + \Bi \ti\leq \bi\, , \\
-\infty &\text{ otherwise}.
\end{cases}
\]
\end{defi}
This general model includes the following well-studied examples:
\begin{itemize} 
\setlength\itemsep{0em}
\item Matching markets \cite{AlaeiKT17,HyllandZ79,vazirani2021computational} in the form 
$u_i(x_i) = \sum_j a_{ij} x_{ij}$ if $\sum_j x_{ij} = 1$, and $-\infty$ otherwise.
\item PLC utility functions (see e.g., \cite{DevanurK08,GargMV16,GargMVY17}) $u_i(x_i)=\min_\ell \{ \sum_j \aijl x_{ij} + \bil\}$  can be modeled as $u_i(x_i) = \max t \text{ s.t. } t \le \sum_j \aijl x_{ij} + \bil,\ \forall \ell$. This includes Leontief utilities as a special case. 
\item Markets with satiation, where an agent may either have the maximum utility limit~\cite{ColeDGJMVY17,BeiGHM19} or consumption constraints~\cite{Mascolell82}, which can be easily captured through constrained PLC functions in most cases.
\end{itemize}
For the special case of PLC utilities, although the problem is PPAD-hard~\cite{ChenT09}\footnote{We note that even the subcases of PLC such as separable PLC is already PPAD-hard for both Fisher~\cite{ChenT09} and Arrow-Debreu models~\cite{ChenDDT09}, and Leontief is PPAD-hard for the Arrow-Debreu model~\cite{CodenottiSVY06}.}, Devanur and Kannan~\cite{DevanurK08} and Kakade, Kearns, and Ortiz~\cite{KakadeKO04} gave polynomial-time algorithms for computing exact and approximate equilibria, respectively, when the number of items is a constant. However, the other significant case of constantly many agents turns out to be much more challenging. In~\cite{DevanurK08} an algorithm is given for fixed number of agents with \emph{separable} PLC utilities, but the case with non-separable PLC utilities remained open. Moreover, apart from theoretical interest, designing simpler and faster algorithms for these cases is crucial for their applications.

\subsection{Our contributions}\label{sec:overview}
 Our main result is a fixed parameter approximation scheme for computing an approximate equilibrium in Fisher model under constrained PLC utilities, where the parameters are the number of agents and the approximation accuracy. The main technical ideas are to use the stronger concept of \emph{thrifty} equilibria and to approximate the input utility functions by \emph{robust} utilities that have favorable marginal properties.  

Before reviewing our algorithm for fixed number of agents, let us start with an easier algorithm for fixed number of items. In this case, a fairly simple grid search works over all possible price combinations with a small stepsize.  This is applicable to the  even more general class of \emph{regular} concave utilities (Theorem~\ref{thm::fisher-fix-item}).  For each price combination, we compute the maximum utility of each agent at these prices, and check whether these utilities can be approximately attained by a feasible allocation also respecting the budget constraints. The existence of an equilibrium guarantees that we find a suitable solution for at least one price combination. This is similar to the grid search approaches used  in \cite[Appendix B]{HyllandZ79} for matching markets, and for other markets in, e.g.,~\cite{Deng2002,KakadeKO04,Papadimitriou2000}. 

The natural starting point for fixed number of agents is to perform a grid search over all possible combinations of utility values with a small stepsize. However, even after fixing the desired utility values for each agent, we need to find both allocations  and prices, a significantly more challenging task. Our approach is to {\em (I)} first find an allocation of the items that meet the utility requirements of each agent, and then {\em (II)} compute prices for which these allocations form a market equilibrium.

Consider an equilibrium with allocations $x^*=\{x^*_{ij}\}_{i,j}$, prices $p^*=\{p^*_j\}_j$ and utility values $u^*=\{u^*_i\}_i$.
For such a two-stage grid search approach to work, a necessary requirement  is that given approximate utility values $u^*_i-\delta\le \tilde u_i\le u_i^*$, $(x,p^*)$ must form an approximate market equilibrium for every allocation $x$ such that  $u_i(x_i)\ge \tilde u_i$ for each agent $i$. This is not true for arbitrary utility functions: not only that $x$ may be very far from $x^*$, but more importantly, the approximate utility value $\tilde u_i$ could be obtained by paying much less than $p^\top x_i^*$.

To address this problem, we make further assumptions both on the utility functions $u_i$ as well as on the equilibrium $(x^*,p^*)$. We require \emph{robust} utility functions, where the change in the utility value is bounded by the change of the budget in a certain critical range of budgets. We then show that every constrained PLC utility function $u_i$ can be approximated by a $\xi$-robust constrained-PLC utility $u_i^\xi$ for any $\xi>0$. We run the algorithm for $u^\xi_i$; the resulting approximate equilibrium will also be an approximate equilibrium for the original $u_i$ with a slightly worse accuracy. The construction of the $u^\xi_i$'s relies on using  \emph{perspective functions} of the $u_i$'s. 

Robustness on its own however does not suffice. A curious phenomenon for constrained utilities is that an agent may not need to spend their entire budget to obtain their most preferred bundle. For example, if the most favored item has price less than 1 in a matching market, the optimal choice of the agent is to purchase the full unit of this item. We will need to require that $(x^*,p^*)$ is a \emph{thrifty} equilibrium: the agents do not only get their most preferred bundle of items, but purchase such a bundle at the cheapest possible costs. (In the matching market example, if there is a tie among most preferred items, all priced less than 1, the agent is only allowed to purchase the cheapest one.)
Fortunately, thriftiness can always be assumed (Theorem~\ref{thm::fisher-exists}):  we show that a thrifty equilibrium always exists in the Fisher model for regular concave utilities mentioned above. The proof uses Kakutani's fixed point theorem. To the extent of our knowledge, existence of (even a non-thrifty) equilibrium is not implied by previous results for constrained PLC utilities.\footnote{We note that~\cite{VaziraniY11} shows NP-hardness of checking equilibrium existence in Fisher model under separable PLC utilities, which seems to require two conditions: first, the sum of prices must equal the sum of budgets, and second, they implicitly assume that the agents are thrifty. Hence, there is no contradiction.} 

Let us now describe the algorithm for robust utilities. In stage {\em (I)}, we can find allocations delivering the utility guesses by solving a linear program. In stage {\em (II)}, the goal is to find prices $p$ that form an approximate equilibrium with $x$. The most challenging part is to ensure that the maximum utility profile available at $p$ is close to the guess $\tilde u_i$ for each agent $i$. This is achieved by considering the dual of the utility maximizing linear program, and applying a variable transformation. After these reductions, suitable prices can be found by  linear programming (Theorem~\ref{lem::fisher-algorithm-const-agent}).

In the above algorithm, we assume that empty allocation is feasible, i.e., $u_i(0) = 0$. However, this is no longer true in matching markets where $u_i(0) = -\infty$, and hence the above algorithm does not directly apply here. We proceed with the natural approach by relaxing the matching constraints to $\sum_j x_{ij} \le 1$ for every $i$. For this relaxation to work, we need to add the requirement on both exact and approximate equilibria that the minimum price is $0$. This can be ensured by exploiting a natural price transformation in the problem.
  We show that this approach works even for a more general model of PLC matching markets with $u_i(x_i) = \min_{\ell} \{\sum_j \aijl x_{ij} + \bil\}$ if $\sum_j x_{ij} = 1$ and $-\infty$ otherwise (Theorems~\ref{thm::plc-matching-items} and \ref{thm::plc-matching-agents}).

The papers~\cite{DevanurK08} and~\cite{AlaeiKT17} give polynomial-time algorithms for computing exact equilibria for the special cases mentioned earlier using a \emph{cell decomposition} technique. Note that in both PLC and matching markets, it is possible that all equilibria are irrational~\cite{Eaves76,vazirani2021computational}; exact equilibria in these works are represented as roots of polynomials.
The cell decomposition arguments partition the parameter space by polynomial surfaces such that in each cell it is easy to decide whether a solution in the particular configuration exists; the number of cells can be bounded using results from algebraic geometry. While the number of cells is polynomial, the results for fixed number of agents (for separable PLC in \cite{DevanurK08} and for matching markets in \cite{AlaeiKT17}) require solving $m^{\mathrm{poly}(n)}$ subproblems and thus may not be practical. In contrast, our algorithm is a fixed parameter scheme in $n$ and the accuracy $\varepsilon$; we need to solve $O((n/\varepsilon^2)^n)$ polynomial-size linear programs.\footnote{We note that $\varepsilon\le 1/(cn)$ is needed for a meaningful approximate equilibrium.} Hence, the complexity of finding an approximate equilibrium is much lower.

For matching markets, we also show that the set of equilibria is non-convex by a simple example of 3 agents and 3 items with tri-valued utility values $a_{ij} \in \{0, 1, 2\}$. To the best of our knowledge, this is the first proof of the non-convexity of equilibria in matching markets.\footnote{We note that~\cite{VaziraniY20-arxiv} presents an example to show non-convexity of equilibria in matching markets. However, their latest version~\cite{vazirani2021computational} does not contain that example, which seems to have only one equilibrium.} Moreover, our example is the simplest one can hope for as for both the bi-valued utility values and two agents case, the set of equilibria is convex; see e.g.,~\cite{GargTV20}. 

Finally, we show that our algorithms also extend to the more general case of the Arrow-Debreu model under PLC utilities (Theorems~\ref{thm::ad-pp} and \ref{thm::ad-fixed-items}). The additional challenge here is to handle budgets that now depend on the prices. For both cases of fixed number of agents and fixed number of goods, we approximate the utilities by robust utilities. This can be done in a simpler way using the special form of the utilities, and in particular it guarantees a lower bound on the minimum price.

\subsection{Related work}\label{sec:rw}
Market equilibrium is an intensely studied concept with a variety of applications, so we briefly mention further relevant results. 
For the classical Fisher model, polynomial-time algorithms are obtained when agents have linear~\cite{DevanurPSV08,Orlin10,Vegh16}, weak gross substitutes~\cite{CodenottiPV05}, and homogeneous utility functions~\cite{Eisenberg61}. For separable PLC utilities, the problem is PPAD-hard~\cite{ChenT09}. 

For the constrained Fisher model, the most famous problem is the Hylland-Zeckhauser scheme for the one-sided matching markets, for which \cite{HyllandZ79} shows the existence of equilibrium, which is recently simplified~\cite{Braverman21}. For matching markets, polynomial-time algorithms are obtained for special cases of constantly many agents (or items)~\cite{AlaeiKT17} and dichotomous utilities~\cite{vazirani2021computational}. Settling its exact complexity is currently open. 

Very recently,~\cite{JalotaPQY21} considers Fisher markets with additional linear constraints, which includes matching markets but not the PLC utilities studied in this paper. It gives a simple fixed-point iterative scheme that converges to an equilibrium in numerical experiments, among other structural results. In particular, it provides a non-convexity example with additional linear utilities, which we note is not a matching market example. 

For the classical Arrow-Debreu model, polynomial-time algorithms are obtained when agents have linear~\cite{DuanGM16,DuanM15,GargV19,Jain07,Ye08} and weak gross substitutes~\cite{CodenottiPV05,BeiGH19,GargHV21} utilities, and beyond that, the problem is essentially PPAD-hard~\cite{ChenDDT09,ChenPY17,CodenottiSVY06,GargMVY17}. 

For the constrained Arrow-Debreu model, an exact equilibrium may not exist even in the case of matching markets~\cite{HyllandZ79}. For this,~\cite{GargTV20} gives the existence of an approximate equilibrium and a polynomial-time algorithm for computing it under dichotomous utilities. 

\paragraph{Overview} The rest of the paper is organized as follows. Section~\ref{sec:model} defines all models and definitions. Section~\ref{sec:Fisher-items} presents an algorithm for computing an approximate Fisher equilibrium under regular concave utilities for a fixed number of items. Section~\ref{sec:Fisher-agents} gives an algorithm for computing an approximate Fisher equilibrium under constrained PLC utilities for a fixed number of agents. Section~\ref{sec:mm} extends algorithms to PLC matching markets and presents an example showing the non-convexity of equilibria. Section~\ref{sec:ad} extends algorithms to the Arrow-Debreu market model. Finally, Section~\ref{sec::proof-existence-fisher} shows the existence of thrifty Fisher equilibrium under regular concave utilities.

\section{Models and definitions}\label{sec:model}

Consider a market with $n$ agents and $m$ divisible items. We assume without loss of generality that there is a unit supply of each item.
Each agent $i$ has a concave utility function $u_i:\R_+^m\to \R\cup\{-\infty\}$.
\begin{defi}
We say that the utility function $u_i:\R_+^m\to \R\cup\{-\infty\}$ is \emph{regular}, if
\begin{itemize}
    \setlength\itemsep{0em}
    \item The function $u_i$ is concave and the domain $K_i=\{x_i\in\R^m_+:\, u_i(x_i)>-\infty\}$ is closed.
    \item $u_i$ restricted to $K_i$ is Lipschitz continuous, i.e. $|u_i(x_i)-u_i(y_i)|\le L\|x_i-y_i\|_2$ for $x_i,y_i\in K_i$. 
    \item $u_i(0)=0$.
\end{itemize}
\end{defi}
We assume that the Lipschitz constant $L$ is the same for all utility functions and is known a priori. This will be relevant
 for the computational complexity of (approximately) solving convex programs with objective $u_i$.
The main requirement in the assumption $u_i(0)=0$  is that $0\in K_i$, i.e., the empty allocation is feasible. If that holds, we can shift the utility function to $u_i(0)=0$. Our main focus will be on the constrained PLC utilities defined in the introduction.

\begin{lemma}\label{lem:CPLC-regular}
Every constrained PLC utility function $u_i$ (Definition~\ref{def:cplc}) with $u_i(0)=0$ is regular. 
For the Lipschitz parameter $L$, $\log L$ is polynomially bounded in the bit-complexity of the input.
\end{lemma}
\begin{proof}
The first property is immediate, and the Lipschitz bound follows by \cite[Corollary 3.2a and Theorem 10.5]{schrijver1998theory}. 
\end{proof}

For prices $p\in\R^m_+$ and a budget $w_i$, we define the optimal utility value
\[
V_i(p,w_i)=\max_{x_i\in\R_+^m}\left\{u_i(x_i):\, p^\top x_i\le w_i\right\}\ ,   
\]
and the \emph{demand correspondence} as the set of utility maximizing bundles that can be purchased at the given budget:
\[
D_i(p,w_i)=\arg\max_{x_i\in\R^m_+}\left\{u_i(x_i):\, p^\top x_i\le w_i\right\}\, ,
\]
Let 
\[
\Vm_i=\max_{ x_i\in [0,1]^m} u_i(x_i)\, .
\] 
be the maximum utility value achievable by purchasing at most 1 unit from each item. Clearly, $V^{\max}_i\ge u_i(0)=0$. Throughout, we make the following normalization assumption:
\begin{equation}\label{eq:V-max}
V^{\max}_i\le 1 \quad\mbox{for each agent }i.
\end{equation}
\hide{If $V^{\max}_i>0$, this can be guaranteed by scaling (we give an interval instead of setting $V^{\max}_i=1$, since one can only approximately compute the value in general).} \hide{If  $V^{\max}_i=0$, then by concavity we must have $u_i(x_i)\le 0$ for all $x_i\in\R^m_+$. We can remove such agents, as they can always be allocated $x_i=0$ at equilibrium.}

 Let 
\[
\C_i(p,w_i)=\min_{x_i\in\R^m_+}\left\{p^\top x_i:\, x_i\in D_i(p,w_i)  \right\}\, 
\]
be the minimum cost of an optimal bundle; we call this the \emph{thrifty cost}. If the market satisfies non-satiation, then $C_i(p,w_i)=w_i$, but it can be strictly less otherwise. Note that if $\C_i(p,w_i)<w_i$, then $V_i(p,w_i)=\max_{ x_i\in \R^m_+} u_i(x_i)$.
We define the  \emph{thrifty demand correspondence} as the set of cheapest optimal bundles.
\[
D_i^t (p,w_i)=\arg\min_{x_i\in\R^m_+}\left\{p^\top x_i:\, x_i\in D_i(p,w_i)  \right\}\, .
\]
Finally, we let 
\[
\Cm_i(p)=\min_{x_i\in\R^m_+}\left\{p^\top x_i:\, x_i\in \R^m, u_i(x_i)\ge \Vm_i  \right\}\, 
\]
denote the minimum cost to achieve $\Vm_i$ at prices $p$. Note that we also allow bundles  here that are not in $[0,1]^m$, i.e., may use more than one unit of an item.

\subsection{The Fisher market model}
In the Fisher market model, we are given $n$ agents and $m$ divisible items of unit supply each. Each agent has a budget $\budgeti$ and a regular utility function $u_i:\R_+^m\to \R\cup\{-\infty\}$. We assume \begin{equation}\label{eq:V-max-pos}
V^{\max}_i > 0 \quad\mbox{for each agent }i.
\end{equation}
\hide{If $V^{\max}_i>0$, this can be guaranteed by scaling (we give an interval instead of setting $V^{\max}_i=1$, since one can only approximately compute the value in general).} If  $V^{\max}_i=0$, then by concavity we must have $u_i(x_i)\le 0$ for all $x_i\in\R^m_+$. We can remove such agents, as they can always be allocated $x_i=0$ at equilibrium.

\begin{defi}[Fisher equilibrium] In a Fisher market with utilities $\{u_i\}_i$ and budgets $\{\budgeti\}_i$, the allocations and prices $(\{x_{i}\}_{i}, \{p_j\}_j)$ form a \emph{market equilibrium} if 
\begin{itemize}
    \setlength\itemsep{0em}
    \item $x_i \in D_i(p,\budgeti)$ for each agent $i$, i.e., each agent buys an optimal bundle subject to budget constraint;
    \item the market clears, i.e., $\sum_i x_{ij} \leq 1$, and $\sum_i x_{ij} = 1$ if $p_j > 0$ for every item $j$.
\end{itemize}
 Further, $(\{x_{i}\}_{i}, \{p_j\}_j)$ is a \emph{thrifty market equilibrium} if we require the stronger $x_i \in D^t_i(p,\budgeti)$ for each agent $i$.
\end{defi}

In Section~\ref{sec::proof-existence-fisher} we prove the following theorem, which shows that regular utilities suffice for the existence of an equilibrium.
\begin{theorem} \label{thm::fisher-exists}
If all agents' utility functions are regular, then a thrifty market equilibrium always exists.
\end{theorem}

\begin{defi}[approximate Fisher equilibrium] \label{def:approx-eq}
In a Fisher market with utilities $\{u_i\}_i$ and budgets $\{\budgeti\}_i$ that satisfies assumption \eqref{eq:V-max}, the allocations and prices 
$(\{x_{i}\}_{i}, \{p_j\}_j)$ form a {\em $(\sigma,\lambda)$-approximate market equilibrium} if 
\begin{itemize}
    \setlength\itemsep{0em}
    \item $u_i(x_i)\ge V_i(p,\budgeti)-\lambda$ ;
    \item $ p^\top x_i \leq \budgeti + \sigma  \sum_i \budgeti$;
    \item $\sum_i x_{ij} \leq 1$, and $\sum_j p_j(1 - \sum_i x_{ij}) \leq  \sigma \sum_i \budgeti$.
\end{itemize}
Similarly, a {\em$(\sigma,\lambda)$-approximate thrifty market equilibrium} satisfies $ p^\top x_i \leq \C_i(p,\budgeti) + \sigma  \sum_i \budgeti$ instead of the second constraint.
A $(\sigma,\sigma)$-approximate (thrifty) market equilibrium will be also referred to as a {\em $\sigma$-approximate (thrifty) market equilibrium}.
\end{defi}

Note that, in order to get a meaningful approximate equilibrium solution, one needs to select $\sigma<1/(cn)$ for some constant $c$, since the error term is $\sigma \sum_i w_i$.

\subsection{The Arrow-Debreu market model}
In the Arrow--Debreu (AD) market model, agents do not have fixed budgets $\budgeti$ but obtain their budget from their shares in the items sold on the market.  Each agent $i$ has an endowment, $\edwi = \{\edw_{ij}\}_j$, 
where $\edwij$ denotes agent $i$'s endowment of item $j$. We assume unit supply of each item, i.e., $\sum_i \edwij = 1$ for every $j$. At prices $p$, each agent $i$ has $\budgeti:=p^\top e_i$ money to spend.

\begin{defi}[AD equilibrium] In an AD market with utilities $\{u_i\}_i$ and endowments $\{e_i\}_i$, the allocations and prices $(\{x_{i}\}_{i}, \{p_j\}_j)$ form a market equilibrium if 
\begin{itemize}
    \setlength\itemsep{0em}
    \item $x_i \in D_i(p, w_i)$;
    \item $\sum_i x_{ij} \leq 1$, and if $p_j > 0$, then $\sum_i x_{ij} = 1$.
\end{itemize}
\end{defi}
Note that scaling the prices $p$ in a market equilibrium by any positive factor results in another equilibrium. Therefore, in the definition of an approximate market equilibrium, we include the normalization constraint $\sum_j p_j \geq 1$.

\begin{defi}[approximate AD equilibrium] In an AD market with utilities $\{u_i\}_i$ and endowments $\{e_i\}_i$ that satisfies assumption \eqref{eq:V-max}, the allocations and prices $(\{x_{i}\}_{i}, \{p_j\}_j)$ form a $(\sigma, \lambda)$-approximate market equilibrium if 
\begin{itemize}
    \setlength\itemsep{0em}
    \item $u_i(x_i) \geq V_i(p, w_i) - \lambda$;
    \item $p^\top x_i \leq w_i + \sigma$;
    \item $\sum_i x_{ij} \leq 1$, and $\sum_j p_j (1 - \sum_i x_{ij}) \leq \sigma $;
    \item $\sum_j p_j \geq 1$.
\end{itemize}
\end{defi}

In contrast to Fisher markets, equilibrium existence in AD markets is guaranteed only under certain assumptions. In fact, already for separable PLC utilities and arbitrary endowments, it is NP-hard to decide whether an equilibrium exists~\cite{VaziraniY11}. 

Correspondingly, our results on Arrow--Debreu markets apply only for the more restrictive piecewise-linear concave (PLC) utilities, and under the assumption that every agent has a positive share in every item. Our assumptions are the following:
\begin{equation}
\label{AD-assumptions}
\begin{aligned}
u_i(x_i) &= \min_l \Big\{\sum_j \aijl x_{ij} + \bil\Big\},\mbox{ and }\min_l\bil=0 \quad \quad \forall i\, ,\qquad \quad \aijl\geq 0 \quad\forall i,j,l\\
 \edwij &> 0 \quad \forall i, j \enspace .
\end{aligned}
\end{equation}

Any PLC utility can be affinely transformed to the above form; hence, $\min_l b_i^l=0$ \hide{and $\aijl\le 1$} is without loss of generality.
Note that $\min_l b_i^l=0$ guarantees that $u_i(0)=0$; thus, these utilities are regular.

Under the even stronger assumption that all $\aijl$'s are positive, existence follows from the classical theorem of Arrow and Debreu. Note that under this assumption, all equilibria are also thrifty.
\begin{theorem}[market equilibrium existence \cite{arrow1954existence}] \label{thm::exist-ad}
If the utilities satisfy \eqref{AD-assumptions} and further all coefficients  $\aijl > 0$, then a market equilibrium always exists. 
\end{theorem}

\section{Approximate Fisher equilibrium for fixed number of items}\label{sec:Fisher-items}
As a warm-up,
we give a simple algorithm for finding an $\varepsilon$-approximate thrifty market equilibrium in Fisher markets for a fixed number of items. The algorithm amounts to approximately solving $O(n\left(\frac{m}{\varepsilon}\right)^m)$ convex programs. 
This is similar to the grid search approaches used, e.g.,~in \cite{Deng2002,HyllandZ79,KakadeKO04,Papadimitriou2000}.

We assume that the utility functions $u_i(x_i)$ are represented by value oracles. For given prices $p$, the maximum utility  $V_i(p,\budgeti)$ and the thrifty cost $\C_i(p,\budgeti)$
 can be obtained as the optimal solution to  convex programs. Using a convex programming algorithm such as the ellipsoid method, we can compute a $\varepsilon$-approximate optimal solutions in oracle-polynomial time in $n$, $m$, the bit-complexity of the vector $p$, $\budgeti$,  $\log L$, and $\log(1/\varepsilon)$ \cite{gls}.

Further, we define the function $F:\R^m_+\to \R$ as the optimal solution to the following convex program in the variables $\{x_{i}\}_i$.
\begin{equation}\label{lp::const-items}
\begin{aligned}
   F(p)=&\min \delta\\
    &u_i(x_i) \geq V_i(p,\budgeti)-\delta, & \forall i \\
    &\sum_i x_{ij} \leq 1, & \forall j  \\
    &\sum_j x_{ij} {p}_j \leq \C_i(p,\budgeti) + \delta \sum_i \budgeti, & \forall i \\
    &\sum_j p_j\left(1-\sum_{i}x_{ij}\right)\le \delta \sum_i \budgeti\\
    &x,\delta \geq 0,
\end{aligned}
\end{equation}
To compute a $\varepsilon$-approximate solution for given prices $p$, we first find $(\varepsilon/2)$-approximate values for $V_i(p,\budgeti)$ for all $i$ and $(\varepsilon\sum_i \budgeti/2)$-approximate values for $\C_i(p,\budgeti)$; then, we again use a convex programming algorithm to find a $(\varepsilon/2)$-approximate solution to the resulting program. 

\begin{lemma}\label{lem:fixed-agent-main} If $F(p)\le \sigma$, then the prices $p$ and allocations $x_i$ give a $\sigma$-approximate thrifty market equilibrium.
 If $(p^*,x^*)$ forms an exact thrifty market equilibrium, and prices $p\in\R^m$ satisfy $p^*_j\le p_j\le p^*_j+\frac{\sigma}{m}\sum_i \budgeti$, then $F(p)\le \sigma$.
\end{lemma}
\begin{proof}
The first claim is immediate by the definition of an approximate thrifty market equilibrium. For the second claim, $F(p^*)=0$ with the optimal solution $x^*$. We show that $(x^*,\sigma)$ is feasible to 
\eqref{lp::const-items}, showing that $F(p)\le \sigma$.
Since $p\ge p^*$, we have $V_i(p,\budgeti)\le V_i(p^*,\budgeti)$, verifying the first constraint. 

To verify the third constraint, we first show that $\C_i(p,\budgeti)\ge \C_i(p^*,\budgeti)$. This is immediate if $\C_i(p,\budgeti)=\budgeti$. If $\C_i(p,\budgeti)<\budgeti$, then $V_i(p,\budgeti)=\max_{x\in \R^m} V_i(x)$, the maximum utility without budget constraint; consequently,  $V_i(p^*,\budgeti)=\max_{x\in \R^m} V_i(x)$. Purchasing such a maximum utility bundle cannot be cheaper at prices $p$, since $p\ge p^*$. Hence,
\[
\sum_j p_j x^*_{ij}\le \sum_j p^*_j x^*_{ij}+\left(\sum_j x^*_{ij}\right)\frac{\sigma}{m}\left(\sum_i \budgeti\right)\le \C_i(p^*,\budgeti)+ \sigma \sum_i \budgeti\le \C_i(p,\budgeti)+ \sigma \sum_i \budgeti\, .
\]
 The last constraint in \eqref{lp::const-items}
 follows by $\sum_j p_j\sum_i x^*_{ij}\ge \sum_j p^*_j\sum_i x^*_{ij}$, and $\sum_j p_j\le \sum_j p_j^*+ \sigma\left(\sum_i \budgeti\right)$.
\end{proof}

\begin{theorem} \label{thm::fisher-fix-item}
Given a Fisher market with $n$ agents, $m$ items, 
and regular concave utility functions $u_i$ given by oracle access, we can compute a $\varepsilon$-approximate thrifty market equilibrium by approximately solving $O(n\left(\frac{m}{\varepsilon}\right)^m)$ convex programs, each in oracle-polynomial time in in $n$, $m$, the bit-complexity of the $\budgeti$'s,  $\log L$, and $\log(1/\varepsilon)$.\footnote{This is essentially $(0, \epsilon)$-approximate equilibrium if we can solve the convex program~\eqref{lp::const-items} exactly; otherwise, we can get $(\lambda, \epsilon)$-approximate equilibrium for arbitrary $\lambda$ at an additional $\log{(1/\lambda)}$ factor.}
\end{theorem}
\begin{proof}
We enumerate all price vectors 
${p}_j = k_j \frac{\varepsilon}{2m} \sum_i \budgeti$ for all integers $k_i$ such that $0 \leq k_j \leq \frac{2m}{\varepsilon} + 1$. For each price vector $p$, we find a $(\varepsilon/2)$-approximate solution to \eqref{lp::const-items}. We output any price vector $p$ for which a solution $(x,\delta)$ with $\delta\le\varepsilon$ is found. Theorem~\ref{thm::fisher-exists} and Lemma~\ref{lem:fixed-agent-main} guarantee the existence of such a solution.
\end{proof}

\section{Approximate Fisher equilibrium for fixed number of agents}\label{sec:Fisher-agents}

In this section, we design a polynomial-time algorithm to compute an approximate equilibrium with constant number of agents for constrained PLC utility functions. 
Recall that $\Vm_i$ is the maximum utility achievable on $[0,1]^m$, and $\Cm_i(p)$ is the minimum cost for achieving utility $\Vm_i$ at prices $p$. Also recall assumptions \eqref{eq:V-max} and \eqref{eq:V-max-pos}, that $\Vm_i\in (0,1]$ for all agents; we make this assumption throughout. The following class of utility functions plays a key role:
\begin{defi}[$\xi$-$\nametwo$ utility function]\label{def::utility-nametwo}
A regular utility function $u_i$ is $\xi$-$\nametwo$, if for any 
bundle $x_i\in \R^m_+$, prices $p\in \R^m_+$ and $\sigma>0$
such that $p^\top x_i\le \Cm_i(p)-\sigma$, there exists a bundle $y\in\R^m_+$ such that $p^\top y\le p^\top x_i+\sigma$ and $u_i(y) \geq u_i(x_i)+\frac{\sigma \xi}{\sum_j p_j}$.
\end{defi}

For this class of utility functions, in Section~\ref{sec::cal-equilibrium-nametwo}, we show how to compute an approximate thrifty market equilibrium. However, not all $\nameone$ utilities are $\xi$-$\nametwo$ for some $\xi > 0$. For example, $u_i(x_i) = 0$ is not $\xi$-$\nametwo$ for any positive $\xi$, while $u_i(x_i) = \varepsilon \sum_j x_{ij}$ 
is $\varepsilon m $-$\nametwo$ for $\varepsilon > 0$. In Section~\ref{sec::approx-nametwo}, we show how to approximate any $\nameone$ utilities by $\nametwo$ utilities. 

In the overall algorithm, we approximate the $\nameone$ utility functions by $\xi$-$\nametwo$ utility functions. Then, we can calculate an approximate thrifty market equilibrium for these $\xi$-$\nametwo$ utility functions. Finally, we show that the approximate market equilibrium we calculated will also be the approximate market equilibrium for the original utility functions (but not necessarily a thrifty one).

\subsection[{Approximate equilibrium for xi-robust utilities}]{Approximate equilibria for $\xi$-$\nametwo$ utilities}\label{sec::cal-equilibrium-nametwo}
By Theorem~\ref{thm::fisher-exists}, we know there exists a thrifty market equilibrium for $\xi$-$\nametwo$ utilities. Let $(x^*,p^*)$ denote a thrifty market equilibrium and let $u^*_i$ denote the utility achieved by agent $i$ at the equilibrium. 

The algorithm has two steps: first, guess each agent's utility at equilibrium and compute a feasible allocation giving each agent at least the guessed utility, and second, compute the prices that, together with the calculated allocation, give an approximate market equilibrium. In Theorem~\ref{thm::fisher-market-constant-item-first}, we give a $(n\delta/\xi,2\delta)$-approximate thrifty equilibrium for some parameter $\delta>0$. This is achieved by solving $O(1/\delta^n)$ 
convex programs.

\subsubsection{Guessing utilities and computing the allocation}
 We first guess agents' utilities at equilibrium by enumerating all possible utilities of each agent, $\tilde{u}_i = k_i \delta$, for $0 \leq k_i \leq \lceil \frac{1}{\delta}\rceil + 1$. Then, we compute a feasible allocation $x=(x_1,\dots,x_n)$ giving $\tilde{u}_i$ utility to agent $i$ using the following program:

\begin{equation}\label{lp::cal-allocation}
\begin{aligned}
& u_i(x_i) \geq \tilde{u}_i\, , &\forall i \\
&\sum_i x_{ij} \leq 1\, , &\forall j \\
& x \geq 0  &
\end{aligned}
\end{equation}
If \eqref{lp::cal-allocation} is infeasible, we move to the next utility profile.
The following lemma shows that if we have a right guess on the utilities, then, with the equilibrium price $p^*$, the spending of agent $i$ at $x_i$ should be similar to that at $x_i^*$. 
\begin{lemma} \label{lem::budget-lower} 
Assume the utility functions are $\xi$-$\nametwo$, $u^*_i - \delta < \tilde{u}_i \leq u^*_i$, and $\{x_i\}_i$ is a feasible solution to \eqref{lp::cal-allocation}. Then,
\begin{enumerate}[(i)]
\item  $\displaystyle{(p^*)^\top x^*_i - \frac{\delta \sum_i \budgeti}{\xi} \leq (p^*)^\top x_i  \leq (p^*)^\top x^*_i + \frac{\delta n \sum_i \budgeti}{\xi}}$, and
\item $\displaystyle{\sum_j p_j^* \left(1 - \sum_i x_{ij}\right) \leq \frac{n\delta \sum_i \budgeti}{\xi}} $.
\end{enumerate}
\end{lemma}
\begin{proof}
We first consider the lower bound in {\em (i)}. Let us denote $\sigma= {\delta \sum_i \budgeti}/{\xi}$.  
For a contradiction, assume $(p^*)^\top x^*_i -\sigma>(p^*)^\top x_i$.
Since this is a thrifty equilibrium,
$(p^*)^\top x^*_i= \C_i(p^*,w_i)\le \Cm_i(p^*)$, since the optimal bundle $x_i^*$ in $[0,1]^m$.

By the $\xi$-robustness property, there exists a bundle $y$ such that 
$(p^*)^\top y\le (p^*)^\top x_i+\sigma<(p^*)^\top x^*_i\le w_i$ and 
\[
u_i(y)\ge u_i(x_i)+\frac{\sigma\xi}{\sum_j p^*_j}\ge u_i(x_i) +\frac{\delta \sum_i \budgeti}{\sum_j p^*_j}> u_i^*(x_i)-\delta+\delta= u_i^*(x_i)\, ,
\]
a contradiction since $u_i^*(x_i)$ is the maximum utility at budget $w_i$. The third inequality uses that all items with $p^*_j>0$ are fully sold, and therefore the sum of the prices is at most the sum of the budgets.

For the upper bound, note that $p^*_j = 0$ if $\sum_i x^*_{ij} < 1$. Therefore, $\sum_j p^*_j \left(\sum_i x^*_{ij}\right) \geq \sum_j p^*_j \left(\sum_i x_{ij}\right)$. Additionally, the lower bound give  $\sum_j p^*_j x_{ij} \geq \sum_j p^*_j x^*_{ij} - \frac{\delta \sum_i \budgeti}{\xi}$ for each $i$, which completes the proof.
Part {\em (ii)} is immediate by summing up the lower bounds in part {\em (i)} for all $i$.
\end{proof}

\subsubsection{Computing the prices}
Assume that for the guesses $\{\tilde{u}_i\}_i$, we found an allocation $x$ that satisfies  \eqref{lp::cal-allocation}, i.e., and allocation of the items that provides at least $\tilde{u}_i$ amount of utility to each $i$. In what follows, our goal is to find prices $p$ that form an approximate equilibrium with $x$.
This is the most challenging part of the algorithm. The prices have to satisfy the following three conditions.

\paragraph{First condition: utility upper bound}
If the guesses  $\{\tilde{u}_i\}_i$ were approximately correct, then 
the maximum utility $V_i(p,w_i)$ achievable at prices $p$ should be close to  $\tilde{u}_i$ for each agent $i$.

Since the utility functions are constrained PLC, we can compute $V_i(p,w_i)$ as
\begin{equation*}
\begin{aligned}
    &\max& \qi^\top z_i + \si^\top t_i\\
    &\text{ s.t. }&  \Ai z_i + \Bi \ti\leq \bi \\
    && p^\top z_i\leq \budgeti\\
    && z_i\ge 0 
\end{aligned}
\end{equation*}
The dual of this program is as follows, using variables $\gamma_i$ and $\beta_i$ for the first two constraints, respectively.   

\begin{equation}\label{dual-lp::utility-upper}
\begin{aligned}
    &\min& \bi^\top\ci + \budgeti \beta_i \\
    &\text{ s.t. }& \Ai^{\top} \ci + \beta_i p \geq \qi \\
    && \Bi^\top \ci = \si \enspace \\
    && \gamma_i,\beta_i\ge 0
\end{aligned}
\end{equation}
For every feasible dual solution,  the objective value provides an upper bound on the optimal utility agent $i$ can get. Therefore, $V_i(p,w_i)\le\tilde{u}_i + \delta$ if and only if there exists a feasible solution $(\ci, \beta_i)$ to \eqref{dual-lp::utility-upper} such that $\ci^\top \bi + \beta_i \budgeti\le\tilde{u}_i + \delta$. 

However, if we also consider the prices $p$ as variables, the program is not linear anymore. For this reason, we use a variable substitution, by letting
$$\frac{1}{\beta_i} \triangleq \overline{\beta_i} \ \ \text{ and  } \ \ \frac{\ci}{\beta_i} \triangleq \overline{\ci}$$ be the variables and we set a lower bound on $\frac{1}{\beta_i}$ such that the optimal solution of \eqref{dual-lp::utility-upper} doesn't change much.

\begin{lemma}\label{lem::utility-upper-used}
For $\delta\in (0,1)$, consider a feasible solution $(\overline{\ci}, \overline{\beta_i}, p)$ to following program,
\begin{equation}\label{dual-lp::utility-upper-used}
\begin{aligned}
    &\bi^\top \overline{\ci}+\budgeti \leq \overline{\beta_i} (\tilde{u}_i + 2\delta)\\
    & {\Ai}^\top \overline{\ci} +  p \geq \overline{\beta_i} \qi \\
    & {\Bi}^\top \overline{\ci} = \overline{\beta_i} \si \\
    &{\budgeti} \leq \overline{\beta_i} 
\end{aligned}
\end{equation}
Then, the optimal utility for agent $i$ to achieve with price $p$ is at most $\tilde{u}_i + 2\delta$. Additionally, if $u^*_i - \delta < \tilde{u}_i \leq u^*_i$, then there exist $\overline{\ci}$ and $\overline{\beta_i}$ such that $(\overline{\ci}, \overline{\beta_i}, p^*)$ is a solution to this program.
\end{lemma}
\begin{proof}
First, if $(\overline{\ci}, \overline{\beta_i}, p)$ is a feasible solution to \eqref{dual-lp::utility-upper-used}, then $(\beta_i = \frac{1}{\overline{\beta_i}}, \ci = \frac{\overline{\ci}}{\overline{\beta_i}}, p)$ is a feasible solution to \eqref{dual-lp::utility-upper}. Therefore, the optimal utility one can get at price $p$ is at most $\bi^\top \ci + \beta_i \budgeti = \frac{1}{\overline{\beta_i}} (\bi^\top\overline{\ci}  +\budgeti) \leq  \tilde{u}_i + 2\delta$.

For the second part, consider the optimal solution $(\ci, \beta_i)$ to \eqref{dual-lp::utility-upper} with the price $p^*$. Then, clearly, $ \bi^\top \ci + \beta_i \budgeti = u_i(x_i^*)$ and $(\ci, \beta_i)$ is also a feasible solution with price $p^*$ and $\budgeti = 0$, which implies $\bi^\top \ci \geq u_i(0) =0$. Combining with the fact that $\bi^\top \ci + \beta_i \budgeti = u_i(x_i^*) \leq 1$, we get $\beta_i \leq \frac{1}{\budgeti}$. Therefore, if we consider the solution $\overline{\ci} = \frac{\ci}{\max\{\beta_i, \frac{\delta}{\budgeti} \}}$, $\overline{\beta_i} = \frac{1}{\max\{\beta_i, \frac{\delta}{\budgeti}\}}$, then it satisfies all conditions in \eqref{dual-lp::utility-upper-used} as $\bi^\top \ci + \beta_i \budgeti = u_i(x_i^*)< \tilde{u}_i + \delta$.
\end{proof}

We note that this is the only part in the algorithm where we rely on the particular form of constrained PLC utilities; all other arguments work more generally, for regular utilities. 

\paragraph{Second condition: budget constraint} The cost of the allocation at prices $p$ must not violate the budget constraints by much:
\begin{align}
    p^\top x_i \leq \budgeti + \frac{n\delta \sum_i \budgeti}{\xi}, \quad\forall i\, . \numberthis \label{lp::budget-constraint-used}
\end{align}
\paragraph{Third condition: market clearing} The market needs to approximately clear:
\begin{align*}
    \sum_j p_j \left( 1 - \sum_i x_{ij} \right) \leq \frac{n\delta \sum_i \budgeti}{\xi}. \numberthis \label{lp::market-clear}
\end{align*}

Lemma~\ref{lem::budget-lower} implies the following: 
\begin{lemma}\label{lem::budget-upper-lp}
Assume that the utility functions are $\xi$-$\nametwo$. If for all agents $i$, ${u}^*_i - \delta < \tilde{u}_i \leq {u}^*_i$, then  for any allocation $x$ for which $\eqref{lp::cal-allocation}$ holds, the optimal prices $p^*$ satisfy \eqref{lp::budget-constraint-used} and \eqref{lp::market-clear}.
\end{lemma}

Note that \eqref{dual-lp::utility-upper-used}, \eqref{lp::budget-constraint-used}, and \eqref{lp::market-clear} are linear in $p$, $\overline{\beta}$, and $\overline{\gamma}$. Combining Lemmas~\ref{lem::utility-upper-used} and \ref{lem::budget-upper-lp}, we have the following:

\begin{theorem} \label{thm::fisher-market-constant-item-first}
Suppose the utility functions are $\xi$-$\nametwo$.  For any $\tilde{u}_i$, $x$ and $p$ such that \eqref{lp::cal-allocation}, \eqref{dual-lp::utility-upper-used}, \eqref{lp::budget-constraint-used}, and \eqref{lp::market-clear} holds,  $(x,p)$ is a $(n\delta/\xi,2\delta)$-approximate market equilibrium.
Additionally, let $(x^*, p^*)$ be any thrifty market equilibrium. If ${u_i}(x^*_i) - \delta < \tilde{u}_i \leq {u}_i(x^*_i)$ for all $i$, then for any $x$ such that $\eqref{lp::cal-allocation}$ holds, $p^*$ is a solution to \eqref{dual-lp::utility-upper-used}, \eqref{lp::budget-constraint-used}, and \eqref{lp::market-clear}.
\end{theorem}

\subsection[Approximating regular by robust utilities]{Approximating $\nameone$ utilities by $\xi$-$\nametwo$ utilities} \label{sec::approx-nametwo}
We introduce an approach to approximate regular utilities by $\xi$-$\nametwo$ utilities. The construction works for the general class of regular functions; for constrained PLC utilities, we show that this operation yields a constrained PLC utility.

Consider a regular utility function $u_i:\R^m_+\to\R\cup\{-\infty\}$, and define the \emph{perspective function} with domain $\R^{m+1}_+$; see \cite[Chapter 5]{Rockafellar}.
\[
\hat u_i(x,\alpha)=\begin{cases}
\alpha u_i\left(\frac{x}{\alpha}\right)\, ,& \text{if } \alpha>0\, ,\\
\lim_{\alpha\to 0} \alpha u_i\left(\frac{x}{\alpha}\right)\, ,&  \text{if }\alpha=0\, .\end{cases}\, 
\]
If $u_i$ is  concave and upper semicontinuous (that hold for regular utilities), then so is $\hat u_i$  \cite[Proposition 2.3(ii)]{Combettes2018}. Also note that $u_i$ is positively homogeneous.
For given $\xi>0$, we define $u_i^\xi:\R^m_+\to\R\cup\{-\infty\}$, where 
\[
\begin{aligned}
u_i^\xi(x_i)=&\max& \hat u_i(x',\alpha)+\hat u_i(x'',1-\alpha)+(1-\alpha)\xi\\
 &\text{ s.t. }& 
 \hat u_i(x'',1-\alpha)\ge (1-\alpha)\Vm_i\\
 &&x'+x''=x_i\\
 &&0\le\alpha\le 1\\
  &&x',x''\ge 0 \enspace .
\end{aligned}
\]
Note that the maximum exists as the objective is convex over a compact domain.
\begin{lemma}\label{lem:perturb-close}
For every regular utility function $u_i$ and $\xi>0$, the following hold for $u_i^\xi$:
\begin{enumerate}[(i)]
\item $u_i(x_i)\le u_i^\xi(x_i)\le u_i(x_i) +\xi$. 
\item 
 We have $\max_{x\in[0,1]^m} u_i^\xi(x)=\Vm+\xi$. For any price vector $p$, the minimum cost of achieving utility $\Vm_i+\xi$ for $u_i^\xi$ is the same as the minimum cost $\Cm_i(p)$ of achieving utility $\Vm_i$ for $u_i$.
\end{enumerate}
\end{lemma}
\begin{proof}
{\bf Part {\em (i)}:} The lower bound  $u_i(x_i)\le u_i^\xi(x_i)$ holds since $x'=x_i$, $x''=0$, $\alpha=1$ is a feasible solution in the definition of $u_i^\xi(x_i)$. The upper bound follows  by the concavity of $\hat u_i$. For any $x'+x''=x_i$ and $\alpha\in [0,1]$, we have
\[
\hat u_i(x',\alpha)+\hat u_i(x'',1-\alpha)+(1-\alpha)\xi\le \hat u_i\left(\frac{x_i}{2},\frac{1}2\right)+(1-\alpha)\xi\le u_i(x_i)+\xi\, .
\]
{\bf\noindent
Part {\em (ii)}:} By the previous part, $\max_{x\in[0,1]^m} u_i^\xi(x)\le\Vm+\xi$. Equality is achieved
for any $x_i\in [0,1]^m$ for which $u_i(x_i)=\Vm_i$ with the choice $x'=0$, $x''=x$, $\alpha=0$. For the second part, consider any price vector $p$. Note that the set $\{x_i\ |\ u_i^\xi(x_i) \geq V_i^* + \xi\}$ is the same as $\{x_i\ |\ u_i(x_i) \geq V_i^*\}$, which completes the proof.
\end{proof}

The next lemma asserts the key property for $\xi$-robustness:
\begin{lemma} \label{lem::utility-perturb-add}
Given prices $p\in\R^m_+$ and $\sigma> 0$, let $x_i\in\R^m_+$ be a bundle such that $p^\top x_i\le \Cm_i(p)-\sigma$. Then, there exists a bundle $y\in\R^m_+$ such that $p^\top y\le p^\top x_i+\sigma$, and
$u_i^\xi(y)\ge u_i^\xi(x_i)+\frac{\sigma \xi}{\sum_j p_j}$.
\end{lemma}
\begin{proof}
Let us use the notation $C=\Cm_i(p)$; by Lemma~\ref{lem:perturb-close}(ii), this is the minimum cost of a bundle of utility $\Vm_i+\xi$.
Let us use the combination $x_i=x'_i+x''_i$ and $\alpha\in[0,1]$ that gives the value of $u_i^\xi(x_i)$, that is,
\[
u_i^\xi(x_i)=\hat u_i(x',\alpha)+\hat u_i(x'',1-\alpha)+(1-\alpha)\xi,
\]
such that $\hat u_i(x'',1-\alpha)\ge (1-\alpha)\Vm_i$. 
By definition, $p^\top x''\ge C$.
Thus, $\alpha p^\top x'+(1-\alpha)C\le p^\top x_i\le C-\sigma$, implying
\[
\alpha\ge \frac{\sigma}{C}\ge \frac{\sigma}{\sum_j p_j}\, ,
\]
where the last inequality uses $C$ is the cost of a bundle in $[0,1]^m$.  
Let $z\in[0,1]^m$ be a bundle such that $u_i(z)=\Vm_i$ and $p^\top z=C$; such a bundle exists since $[0,1]^m$ is a compact domain.
  Let
\[
\beta=\frac{\sigma}{\sum_j p_j}\, ,\quad
y'=\frac{\alpha-\beta}{\alpha}\cdot x'\, ,\quad
y''=x''+\beta z\, ,
\quad \mbox{and}\quad y=y'+y''\, .
\]
Note that 
\[
p^\top y<  p^\top x+\beta p^\top z= p^\top x+ \frac{\sigma C}{\sum_j p_j}\le p^\top x+ \sigma\, ,
\]
satisfying the required bound on the cost. The rest of the proof amounts to showing 
$u_i^\xi(y)\ge u_i^\xi(x_i)+\frac{\sigma \xi}{\sum_j p_j}$.
\begin{claim}\label{cl:ypp-good}
$\hat u_i(y'',1-\alpha+\beta)\ge  \hat u_i(x'',1-\alpha)+\beta\Vm_i\ge (1-\alpha+\beta)\Vm_i$.
\end{claim}
\begin{claimproof}
By the homogeneity and concavity of $\hat u_i$,
\[
\hat u_i(y'',1-\alpha+\beta)=2\hat u_i\left(\frac{y''}{2},\frac{1-\alpha+\beta}2\right)\ge \hat  u_i(x'',1-\alpha)+\hat u_i(\beta z,\beta)\ge \hat u_i(x'',1-\alpha)+\beta\Vm_i\, .
\]
The last inequality in the claim follows by noting that also $\hat u_i(x'',1-\alpha)\ge (1-\alpha)\Vm_i$.
\end{claimproof}
\begin{claim} \label{cl:xp-bad}
$\hat u_i(y',\alpha-\beta)= \frac{\alpha-\beta}{\alpha} \hat u_i(x',\alpha)> \hat u_i(x',\alpha)-\beta \Vm_i$.
\end{claim}
\begin{claimproof}
The first inequality is by definition of $\hat u_i$. The second inequality is equivalent to $\hat u_i\left({x'},\alpha\right)< \alpha\Vm_i$.
Assume for a contradiction $\hat u_i\left({x'},\alpha\right)\ge \alpha\Vm_i$. Then, replacing $x''$ by $x'+x''$ and $\alpha$ by 0 results in a better combination using the concavity of $\hat u_i$ as in the previous claim.
\end{claimproof}
By Claim~\ref{cl:ypp-good}, $y=y'+y''$ is a feasible decomposition in
the definition of $u_i^\xi$ with coefficient $\alpha-\beta$. Further, note that
$\hat u_i(y',\alpha-\beta)= \frac{\alpha-\beta}{\alpha} \hat u_i(x',\alpha)$
We get
\[
\begin{aligned}
u_i^\xi(y)&\ge \hat u_i (y',\alpha-\beta)+ \hat u_i (y'',1-\alpha+\beta)+(1-\alpha+\beta)\xi \\
&\ge \hat u_i (y',\alpha-\beta)+ \hat u_i (x'',1-\alpha)+\beta\Vm_i+(1-\alpha+\beta)\xi\\
&\ge \hat u_i (x',\alpha)+ \hat u_i (x'',1-\alpha)+(1-\alpha+\beta)\xi\\
&\ge u_i^\xi(x_i)+\beta\xi=u_i^\xi(x_i)+\frac{\sigma\xi}{\sum_j p_j}
\, ,
\end{aligned}
\]
as required. The second inequality used Claim~\ref{cl:ypp-good} and the  third inequality used Claim~\ref{cl:xp-bad}.
\end{proof}

Let us now turn to constrained PLC utilities.
\begin{lemma}\label{lem:CPLC-robust}
Let $u_i$ be a constrained PLC utility with $u_i(0)=0$ and $\Vm_i>0$, and $\xi>0$. Then, $u_i^\xi$ is also a constrained PLC utility, and is $\xi$-robust. The bit-length of the LP description of $u_i^\xi$ is polynomial in the LP description of $u_i$ and of $\log\xi$.
\end{lemma}
\begin{proof}
Recall the form of the constrained PLC utility as
\begin{align*}
    u_i(x_i) =
\max_{t_i} \qi^\top x_i + \si^\top t_i  \text{ s.t. }  \Ai x_i + \Bi \ti\leq \bi\, ,
\end{align*}
where the value is $-\infty$ if the problem is infeasible.
It is easy to verify that the following linear program gives an equivalent description of $ u^\xi_i$:
\begin{equation}\label{lp::nametwo-utility-constraint}
\begin{aligned}
u_i^\xi(x)&= \qi^\top x_i + \si^\top t_i  + (1 - \alpha) \xi\\
 &\Ai x'_i + \Bi t'_i \leq \alpha \bi  \\
  &\Ai x''_i + \Bi  t''_i \leq (1 - \alpha) \bi\\
& \qi^\top x''_i + \si^\top t''_i \geq  (1 - \alpha) \Vm_i\\
&x'_i + x''_i=x_i  \\
& t'_i + t''_i =t_i\\
 &x_i',x_i''\ge 0\\
& 1\ge \alpha \ge 0
\end{aligned}
\end{equation}
Hence, $u^\xi_i$ is also constrained PLC. If it is regular, then it is $\xi$-robust by Lemma~\ref{lem::utility-perturb-add}. By Lemma~\ref{lem:CPLC-regular}, regularity only requires $u^\xi_i(0)=0$ in this case. 
This follows by Lemma~\ref{lem:perturb-close}(i) and the assumptions $u_i(0)=0<\Vm_i$. Finally, the statement on bit-complexity follows since $\Vm_i$ is the optimum value of a linear program formed by the LP defining $u_i$ and a box constraint. Therefore, $\Vm_i$ is polynomially bounded in the input. 
\end{proof}

\begin{theorem}
\label{lem::fisher-algorithm-const-agent}
In a Fisher market with $n$ agents, $m$ items and regular constrained PLC utility functions and $\sigma<1$, we can find a $\sigma$-approximate market equilibrium by solving  $O\left(\left(\frac{n}{\sigma^2}\right)^n\right)$ linear programs, each polynomially bounded in the input size.
\end{theorem}
\begin{proof}
Let us set $\delta=\sigma^2/(2n)$ and $\xi=\sigma/2$.
We first replace the utility functions $u_i$ by $ u_i^\xi$ as in Lemma~\ref{lem:CPLC-robust}. Then, we guess all combinations  
$\tilde{u}_i = k_i \delta$, for $0 \leq k_i \leq \lceil \frac{1}{\delta}\rceil + 1$. We calculate the allocations  $x$ as in \eqref{lp::cal-allocation}; if no such allocation exists, we proceed to the next guess. If $x$ is feasible to \eqref{lp::cal-allocation}, then we check if  prices $p$ satisfying \eqref{dual-lp::utility-upper-used}, \eqref{lp::budget-constraint-used}, and \eqref{lp::market-clear} exist.
Theorem~\ref{thm::fisher-market-constant-item-first} guarantees the existence both $x$ and $p$ for at least one choice of the $\tilde u_i$'s. 
This gives a $(n\delta/\xi,2\delta)$-approximate equilibrium for the utilities $u_i^\xi$, and by Lemma~\ref{lem:perturb-close}(i), a 
$(n\delta/\xi,2\delta+\xi)$-approximate equilibrium for the original utilities $u_i$. By the choice of $\delta$ and $\xi$, this is a $\sigma$-approximate equilibrium.
\end{proof}


\section{PLC Matching Markets}\label{sec:mm}
In the Hylland-Zeckhauser matching market equilibrium \cite{HyllandZ79}, agents have unit budgets  and linear utilities with the additional restriction that every agent has to purchase exactly one unit of good. 
We now consider the following generalization with PLC utilities for nonnegative values $\aijl,\bil\ge 0$.
\begin{align}\label{eq:mm-u-i}
        u_i(x_i) = \begin{cases}
    \min_l \left\{\sum_j \aijl x_{ij} + \bil \right\} \, ,     & \quad \text{if } \sum_j x_{ij} = 1\ ,\\
    -\infty & \quad \text{otherwise.}
  \end{cases}
    \end{align}
Throughout this section, we assume $\budgeti=1$ for all agents, as standard in the matching market model. We also assume $n\le m$, i.e., there are at least as many items as agents that is necessary for feasibility.
 We refer to this problem as the \emph{PLC matching market} problem.
 Let 
$\Vmm_i=\max_{x_i\in\R^m_+}u_i(x_i)$ be the maximum achievable utility of agent $i$; the matching constraint guarantees this is finite.
Similarly to \eqref{eq:V-max}, without loss of generality we can apply affine transformations to the utilities so that
\begin{equation}\label{eq:mm-scaling}\Vmm_i\le 1\quad\mbox{and}\quad \min_l \bil = 0\quad\mbox{for each agent }i.
\end{equation}

Note that, even though this utility function is constrained PLC, it is \emph{not regular}: $u_i(0)=-\infty$. For this reason, we cannot directly apply the results in Sections~\ref{sec:Fisher-items} and~\ref{sec:Fisher-agents}. The existence of an equilibrium is also not covered by Theorem~\ref{thm::fisher-exists}.
A key tool to tackle this model is the following price transformation with strong invariance properties that enables us to restrict our attention to (approximate) equilibria where $\min_j p_j=0$.

\begin{lemma}[\cite{vazirani2021computational}]\label{lem:price-transform}
For a PLC matching market model,
let $p\in\R^m_+$, and $r>0$ such that $p'_j=1+r(p_j-1)\ge 0$ for all items $j$. Then, 
 $D_i(p,1)=D_i(p',1)$ for every agent $i$.
Consequently, if there exists a market equilibrium 
$(\{x^*_i\}_i, \{p^*_j\}_j)$, then there exists one with $\min_j p_j=0$.
\end{lemma}
\begin{proof}
Both $D_i(p,1)$ and $D_i(p',1)$ only contain bundles $x_i$ with
$\sum_j x_{ij}=1$.
Since $1-(p')^\top x_i=r(1-p^\top x_i)$ for such a bundle, the price of a bundle satisfies 
  $p^\top x_i\le 1$ if and only if it satisfies $(p')^\top x_i\le 1$. This implies $D_i(p,1)=D_i(p',1)$. For the second part, consider any market equilibrium 
$(\{x^*_i\}_i, \{p^*_j\}_j)$. If there exists a good at price $p^*_j<1$, then we can select the largest $r$ value such that this transformation gives $\min_j p_j'=0$. The first part guarantees that $(\{x^*_i\}_i, \{p'_j\}_j)$ is also a market equilibrium. Otherwise, $p^*_j = 1$ for all $j$. In this case, setting $p_j = 0$ for all $j$ will also be a matching market equilibrium.
\end{proof}

In light of this transformation, we note that the $(\sigma,\lambda)$-approximate (thrifty) equilibrium concept as in Definition~\ref{def:approx-eq} is unsatisfactory. Assume $n=m$, i.e., the number of items is the same as the number of agents. Let  $(\{x_{i}\}_{i}, \{p_j\}_j)$ be a $(\sigma,\lambda)$-approximate equilibrium. Then, for any choice of $0<\sigma'\le \sigma$, we can select $r>0$ such that $(\{x_{i}\}_{i}, \{p'_j\}_j)$ will be a $(\sigma',\lambda)$-approximate (thrifty) equilibrium. This is because $(p')^\top x_i$ becomes arbitrarily close to 1, and the third constraint is satisfied since $\sum_{i} x_{ij}=1$ for all $j$ follows if $n=m$.

In accordance with Lemma~\ref{lem:price-transform}, we will look for approximate (thrifty) equilibria with the additional requirement that $\min_j p_j=0$. In Section~\ref{sec:partial}, we show that approximate equilibrium results can be obtained by reducing to an associated partial matching market. In Section~\ref{sec:non-convexity}, we give a simple counterexample showing that the set of equilibria is non-convex already for the standard matching market model with three agents and three items.

\subsection{From partial to perfect matchings}\label{sec:partial}
Both for showing the existence of equilibria, as well as for the algorithms, we relax the perfect matching requirement $\sum_j x_{ij}=1$ to the \emph{partial} matching constraint
$\sum_j x_{ij}\le1$. That is, for the same parameters $\aijl,\bil$, we let
 \begin{align}\label{eq:mmr-u-i}
        u'_i(x_i) = \begin{cases}
    \min_l \left\{\sum_j \aijl x_{ij} + \bil \right\} \, ,     & \quad \text{if } \sum_j x_{ij} \leq 1\, ,\\
    -\infty & \quad \text{otherwise.}
  \end{cases}
    \end{align}
Using the assumption \eqref{eq:mm-scaling}, $u'_i(0)=0$, and therefore the $u'_i$'s are regular utilities. For a PLC matching market with utilities $u_i$ as in \eqref{eq:mm-u-i}, we will refer to the market that replaces the $u_i$'s by the $u'_i$'s as the associated \emph{PLC partial matching market}.

The next two lemmas show the close relationship between equilibria in these markets. In the proofs, we use $V_i(p,1)$ for the optimal utility for $u_i$ and $C_i(p,1)$ the minimum price of an optimal bundle; we let $V'_i(p,1)$ and $C'_i(p,1)$ denote the same for $u'_i$. Clearly, $V'_i(p,1)\ge V_i(p,1)$.
\begin{lemma}\label{lem::crp-fisher-matching}
Let $(\{x_{i}\}_{i}, \{p_j\}_j)$ be a thrifty PLC matching market equilibrium with $\min_j p_j=0$. Then, $(\{x_{i}\}_{i}, \{p_j\}_j)$  is also a thrifty  market equilibrium in the associated PLC partial matching market.
\end{lemma} 
\begin{proof}
Using that $p_k=0$ for some good $k$,  for every $x'_i$ with $\sum_j x'_{ij}\le 1$ there exists a bundle $\tilde x_i\ge x'_i$ with $\sum_j \tilde x_{ij}= 1$ that has the same cost 
and $u_i(\tilde x_i)\ge u_i'(x'_i)$. Consequently, $V'_i(p,1)= V_i(p,1)$, and by the same token, $C_i(p,1)=C'_i(p,1)$. The statement follows.
\end{proof}

In the other direction, we show that approximate  (thrifty) equilibria in the associated  PLC partial matching market have $\min_j p_j=0$, then this can be extended to the original PLC matching market. This also applies to exact equilibria with $\sigma=\lambda=0$.

\begin{lemma}\label{lem::build-matching-fisher}
For a PLC matching market, consider a $(\sigma, \lambda)$-approximate (thrifty) equilibrium
$(\{x'_i\}_i, \{p'_j\}_j)$ in the associated PLC partial matching market, and assume $\min_j p'_j=0$. Then, in $O(m)$ time  
 we can construct a $(2 \sigma, \lambda)$-approximate (thrifty) matching equilibrium $(\{{x}_i\}_i, \{p'_j\}_j)$ in the original market.
\end{lemma}
\begin{proof}
Given $(\{x'_i\}_i, \{p'_j\}_j)$, we  arbitrarily assign those items which are not fully allocated to those agents such that $\sum_j x'_{ij} < 1$; this can be easily done in $O(m)$ time (recall $m\ge n$). Let $\{{x}_i\}_i$ denote the resulting allocations with $\sum_j x_{ij}=1$.

 Recalling that all $w_i=1$, the approximate equilibrium means  $p^\top x_i\le 1+n\sigma$ and $\sum_j p'_j (1 - \sum_i x'_{ij}) \leq \sigma \sum_i \budgeti=n\sigma$. Hence, the spending for each agent (after assignment) can be at most $2n\sigma+1$. 
As in the previous proof, $\min_j p_j=0$ guarantees that $V_i(p,1)=V_i'(p,1)$ and $C_i(p, \budgeti)=C_i'(p, \budgeti)$.
The utility requirement follows since
$V_i(p,1)-\lambda= V'_i(p,1)-\lambda\le u'_i(x'_i)\le u_i(x_i)$.
Further, if $(\{x'_i\}_i, \{p'_j\}_j)$ was an approximate thrifty market equilibrium, then thriftiness for $(\{{x}_i\}_i, \{p_j\}_j)$ follows since  the spending can only be increased by $n\sigma$.
\end{proof}

We can now derive the existence of an equilibrium, as well as algorithms for approximate equilibria, by making use of the results for PLC partial matchings that are regular utilities.

\begin{theorem} \label{thm::ext-matching}
In every PLC matching market, there exists a thrifty market equilibrium $(\{x_{i}\}_{i}, \{p_j\}_j)$ 
with $\min_j p_j=0$.
\end{theorem}
\begin{proof}
For the $u'_i$ utilities in the associated PLC partial matching market, 
Theorem~\ref{thm::fisher-exists} guarantees the existence of an equilibrium $(\{x'_i\}_i, \{p'_j\}_j)$. If $\min_j p'_j=0$, then  Lemma~\ref{lem::build-matching-fisher} for $\sigma=\lambda=0$ gives an equilibrium in the PLC matching market with the $u_i$'s. If $\min_j p'_j>0$, then all goods must be fully sold, hence $\sum_{i,j} x'_{ij}=m\ge n$. This cannot happen if $m>n$; and if $m=n$ this implies that all agents are getting one unit in $x'$, i.e., $\sum_{j} x'_{ij}=1$ for all $i$. Consequently, $(\{x'_i\}_i, \{p'_j\}_j)$ is already an equilibrium in the PLC matching market. By Lemma~\ref{lem:price-transform}, this can be transformed to one with $\min_j p_j=0$.
\end{proof}

For fixed number of items, we can thus use the algorithm in Section~\ref{sec:Fisher-items} for $u'_i$, and transform it using Lemma~\ref{lem::build-matching-fisher} for $u_i$. In order to find an approximate thrifty equilibrium for the $u'_i$'s with 
$\min_j p_j=0$;  we only enumerate over price combinations where one of the prices is 0. Theorem~\ref{thm::ext-matching} and Lemma~\ref{lem::crp-fisher-matching} guarantee the existence of such a solution.
\begin{theorem}[approximate thrifty PLC matching market equilibrium with fixed number of items]\label{thm::plc-matching-items}
Given a PLC matching market with $n$ agents, $m$ items, 
and PLC utilities $\{u_i\}_i$, we can compute an $\varepsilon$-approximate thrifty PLC matching market equilibrium by solving $O(n\left(\frac{m}{\varepsilon}\right)^m)$ linear programs, each in polynomial time in the input size. 
\end{theorem}
Similarly, for fixed number of agents, we can use the results in Section~\ref{sec:Fisher-agents} for $u'_i$ in conjunction with 
 Lemma~\ref{lem::build-matching-fisher} to compute an approximate PLC matching market equilibrium (but not necessarily a thrifty one). The only modification needed is that we fix the price of some good to $p_j=0$; this results in an additional factor $m$ in the running time.
\begin{theorem}[approximate PLC matching market equilibrium with fixed number of agents]\label{thm::plc-matching-agents}
Given a PLC matching market with $n$ agents, $m$ items, 
and PLC utilities $\{u_i\}_i$, we can compute a $\sigma$-approximate  PLC matching market equilibrium by solving $O\left(m \left( \frac{n}{\sigma^2} \right)^n\right)$ linear programs, each in polynomial time in the input size. 
\end{theorem}
Finally, for the original Hylland-Zeckhauser model with linear utilities, we show that the stronger concept of  an approximate \emph{thrifty}  equilibrium can also be computed, by exploiting the  simpler structure in this case.

    \begin{theorem}[approximate thrifty  matching market equilibrium with fixed number of agents]
    Given a  matching market with $n$ agents, $m$ items, 
and linear utility function 
 \begin{align*}
        u_i(x_i) =  \begin{cases}\sum_j a_{ij} x_{ij}\, ,      & \quad \text{if } \sum_j x_{ij} = 1\, ,\\
    -\infty & \quad \text{otherwise.}
  \end{cases}
    \end{align*}
 we can compute a $\sigma$-approximate thrifty market equilibrium by solving $O\left(m \left( \frac{n}{\sigma^2} \right)^n\right)$ linear programs, each in polynomial time in the input size.
\end{theorem}
\begin{proof}
Similar to the PLC case, we first calculate a thrifty approximate equilibrium for the associated partial matching market such that  $\min_j p_j = 0$ and then transform it into a thrifty approximate matching market equilibrium. 
The transformed  $\xi$-$\nametwo$ utility       ${u'_i}^{\xi}(x_i)$ used in the algorithm (see \eqref{lp::nametwo-utility-constraint})
can be written in the following simpler form. Let $J=\arg\max_j \aij$, and
    \begin{align*}
        {u'_i}^{\xi}(x_i) = \begin{cases}
    \sum_{j\notin J} \aij x_{ij}+\sum_{j\in J}(\aij+\xi)x_{ij} \, ,     & \quad \text{if } \sum_j x_{ij} \leq 1\, , \\
    -\infty & \quad \text{otherwise}.
  \end{cases} \numberthis \label{utility-matching-nametwo}
    \end{align*}

Let us calculate an approximate thrifty market equilibrium $(\{x_{i}\}_{i}, \{p_j\}_j)$ for ${u'_i}^{\xi}$ as in Section~\ref{sec::cal-equilibrium-nametwo} with two slight modifications.

     We first enumerate all possible $\tilde{u}_i$ for $\tilde{u}_i = \delta k_i$ for $0 \leq k_i \leq \lceil \frac{1 + \xi}{\delta} \rceil + 1$ and one $j$ such that $p_j = 0$; then, we calculate $\{x_i\}_i$ by  \eqref{lp::cal-allocation}; and finally, we calculate the price $\{p_j\}_j$. When calculating the price, in addition to  \eqref{dual-lp::utility-upper-used}, \eqref{lp::budget-constraint-used}, and \eqref{lp::market-clear}, we add constraints $p_j = 0$ and 
     \begin{equation}\label{eq:J}
     p^\top x_i \leq p_{j'} + \frac{n^2 \delta}{\xi}
     \quad \quad \forall j' \in J\, .
     \end{equation}
     Recall that $\frac{n^2 \delta}{\xi}=\frac{n \delta}{\xi}\sum_i \budgeti$ by the assumption that all budgets are 1.
This additional last inequality makes the difference compared to the general PLC algorithm. We exploit this in the following claim.
    \begin{claim}
    If $p^\top x_i \leq p_{j'} + \frac{n \delta}{\xi}\sum_i \budgeti$ for  $j' \in J$; and $p^\top x_i \leq 1 + \frac{n^2 \delta}{\xi}$, then $p^\top x_i \leq C_i(p, 1) + \frac{n^2 \delta}{\xi}$. Additionally, let $(\{x^*_i\}_i, \{p^*_j\}_j)$ be any thrifty market equilibrium. If ${u}^*_i - \delta < \tilde{u}_i \leq {u}^*_i$ for all $i$, then for any $\{x_i\}_i$ such that $\eqref{lp::cal-allocation}$ holds, \eqref{eq:J} is also valid for $p^*$.\end{claim}
    \begin{claimproof}
    The first part follows as $C_i(p, 1) = \min\{1, \min_{j \in J} p_j\}$. The second part is true because, for $j' \in J$,
    \begin{align*}
    {p^*}^\top x_i &\leq {p^*}^\top x^*_i + \frac{n^2 \delta }{\xi} ~~~~~\text{(by Lemma~\ref{lem::budget-lower})} \\
    &\leq p^*_{j'} + \frac{n^2 \delta}{\xi}. ~~~~~\text{(as ${p^*}^\top x^*_i = C_i(p^*, 1) \leq p_j'$) } \qedhere
\end{align*}
    \end{claimproof}
    
   Combining this observation with  Theorem~\ref{thm::fisher-market-constant-item-first}, Lemma~\ref{lem::crp-fisher-matching}, and 
   Theorem~\ref{thm::ext-matching}, this procedure will output a $(\delta n/ \xi, \xi + 2\delta)$-approximate thrifty equilibrium for the associated partial matching market.
   
       Finally, by Lemma~\ref{lem::build-matching-fisher}, we construct an approximate matching market equilibrium from the approximate Fisher market equilibrium.
 The theorem follows by choosing $\delta = \sigma^2/(4n)$ and $\xi = \sigma/2$.
\end{proof}

\subsection{Non-convexity example}\label{sec:non-convexity}
In this section, we give a simple example which shows that the sets of allocations and prices are non-convex.  The example consists of three agents, three items and the utilities are linear for these agents: $u_i(x_i) = \sum_j a_{ij} x_{ij}$. Each agent has a budget of $1$ dollar.
\begin{table}[h!]
\centering
\begin{tabular}{|c|c|c|c|}
\cline{1-4}
        & item 1 & item 2 & item 3   \\ \cline{1-4}
agent 1 & 1      & 1      & 2        \\ \cline{1-4}
agent 2 & 0      & 1      & 2        \\ \cline{1-4}
agent 3 & 1      & 1      & 2        \\ \cline{1-4}
\end{tabular}
\caption{Utility matrix ($\aij$)}
\end{table}

Given the utility functions, the following prices and allocations are two of the equilibria of the matching market.
\begin{table}[!htb]
    \begin{minipage}{.5\linewidth}
      
      \centering
        \begin{tabular}{|c|c|c|c|}
\cline{1-4}
        & item 1 & item 2 & item 3   \\ \cline{1-4}
agent 1 & 0.5      & 0      & 0.5        \\ \cline{1-4}
agent 2 & 0      & 1      & 0        \\ \cline{1-4}
agent 3 & 0.5      & 0      & 0.5        \\ \cline{1-4} \hline \hline
price  & 0      &  1     & 2        \\ \cline{1-4}
\end{tabular}
\caption{Price 1 ($p^{(1)}$) and Allocation 1 ($x^{(1)}$)}
    \end{minipage}%
    \begin{minipage}{.5\linewidth}
      \centering
        
        \begin{tabular}{|c|c|c|c|}
\cline{1-4}
        &item 1 & item 2 & item 3   \\ \cline{1-4}
agent 1 & 2/3      & 0      & 1/3        \\ \cline{1-4}
agent 2 & 0      & 2/3      & 1/3        \\ \cline{1-4}
agent 3 & 1/3      & 1/3      & 1/3        \\ \cline{1-4}\hline \hline
price  & 0      &  0     & 3        \\ \cline{1-4}
\end{tabular}
\caption{Price 2 ($p^{(2)}$) and Allocation 2 ($x^{(2)}$)}
    \end{minipage} 
\end{table}

The following two lemmas show that neither the set of allocations nor the set of prices is convex. 
\begin{lemma} $\frac{p^{(1)} + p^{(2)}}{2}$ is not an equilibrium price.
\end{lemma}
\begin{proof}
Note that $\frac{p^{(1)} + p^{(2)}}{2} = (0, 0.5, 2.5)$. In this case, both agent $1$ and agent $3$ will not be interested in item $2$. This implies agent $2$ will get item $2$ fully. However, given the price, agent $2$ will buy some of item $3$, which provides a contradiction.
\end{proof}

\begin{lemma}
$\frac{x^{(1)} +x^{(2)}}{2}$ is not an equilibrium allocation.
\end{lemma}
\begin{proof}
Note that in any equilibrium, the price of item $3$ should be strictly larger than $1$. This implies all agents will spend out all their budgets. Let the price of item $3$ be $1 + \alpha$ for some $\alpha > 0$. Since agent $1$ get $7/12$ of item $1$ and $5/12$ of item $3$, the price of item $1$ is $1 - \frac{5}{7} \alpha$. Similarly, since agent $2$ get $5/6$ of item $2$ and $1/6$ of item $3$, the price of item $2$ is $1 - \frac{1}{5} \alpha$. Since $\alpha > 0$, given the price $(1 - \frac{5}{7} \alpha, 1 - \frac{1}{5} \alpha, 1 + \alpha)$, agent $3$ will not buy item $2$, which contradicts allocation  $\frac{x^{(1)} +x^{(2)}}{2}$.
\end{proof}

\section{Approximate Arrow-Debreu market equilibrium}\label{sec:ad}

In this section, we show how to compute an approximate Arrow-Debreu (AD) market equilibrium with  \eqref{AD-assumptions}, when either the number of agents or items is constant. We note that while an equilibrium is only guaranteed to exists with $\aijl>0$ (Theorem~\ref{thm::exist-ad}), we show the existence of approximate equilibria under the weaker assumption $\aijl\ge0$.

Both for a fixed number of items and fixed number of agents, the key is to perturb the utilities to $\xi$-robust utilities. However, we use a different, simpler perturbation as in \eqref{lp::nametwo-utility-constraint} using that the utilities are PLC.
\begin{equation}\label{eq:bu}
\bu^{\xi}_i(x_i) = \min_l \left\{\sum_j \left(\aijl + \frac{\xi}{m}\right) x_{ij} + \bil\right\}\, .
\end{equation}
\begin{lemma}\label{lem::ad-u-1}
$\bar u_i^{\xi}$ is a $\xi$-$\nametwo$ utility function. Additionally, for any $x_i$, $\bu^{\xi}_i(x_i) \geq u_i(x_i)$, and for any $x_i\in [0,1]^m$, $\bu^{\xi}_i(x_i) \leq u_i(x_i) + \xi$.
\end{lemma}
\begin{proof}
For the $\xi$-robustness, regularity is immediate by assumption \eqref{AD-assumptions}.
The other property in the definition follows since  
with additional $\sigma$ amount of money, agent $i$ can buy each item with amount $\frac{\sigma}{\sum_j p_j}$, which, in total, provides at least $\frac{\sigma \xi}{\sum_j p_j}$ additional amount of utility. The other bounds are immediate.
\end{proof}

\subsection{Fixed number of agents}
Similar to the Fisher market case, we first show how to compute market equilibrium with $\xi$-\nametwo\ utilities, and then we show how to approximate the PLC utilities by $\xi$-\nametwo\ utilities.

For $\xi$-robust utilities, the algorithm is a simple modification of the one in Section~\ref{sec::cal-equilibrium-nametwo}. The main difference is that the budgets are not fixed but defined according to the prices.

We guess the equilibrium utilities (approximately) for all players, $\{\tilde{u}_i\}_i$, and then calculate the allocation by \eqref{lp::cal-allocation}. We use the linear system comprising \eqref{dual-lp::utility-upper-used}, \eqref{lp::budget-constraint-used}, and \eqref{lp::market-clear} and the following additional constraints:
\begin{align*}
    &p^\top \edwi = \budgeti\quad \forall i\, , \\
    &\sum_j p_j = 1\, . \numberthis \label{lp::endowment-budget}
\end{align*}

We can obtain the following similarly to Theorem~\ref{thm::fisher-market-constant-item-first}: 
\begin{theorem} \label{thm::ad-pp}
Suppose the utility functions are $\xi$-\nametwo\ and the  Arrow Debreu market equilibrium $(\{x^*_i\}_i, \{p^*_j\}_j)$ exists. For any $\{\tilde{u}_i\}_i$, $\{x_{i}\}_i$ and $\{p_j\}_j$ such that \eqref{lp::cal-allocation}, \eqref{dual-lp::utility-upper-used}, \eqref{lp::budget-constraint-used}, \eqref{lp::market-clear}, \eqref{lp::endowment-budget} holds, $(\{x_i\}_i, \{p_j\}_j)$ is a $(n\delta/\xi,2 \delta)$-approximate equilibrium.
\hide{\begin{enumerate}
    \item for any $i$, $\tilde{u}_i + \delta + \eta \geq  \max_{x_{ij} \geq 0, p^\top x_i \leq p^\top \edwi} \{ \min \{\sum_j \aijl x_{ij} + \bil\}$;
    \item for any $i$, $u_i(x_i) \geq \tilde{u}_i$;
    \item for any $i$, $ p^\top x_i \leq \budgeti + \frac{n\delta \sum_i \budgeti}{\xi}$;
    \item for any $j$, $\sum_i x_{ij} \leq 1$;
    \item $\sum_j p_j ( 1 - \sum_i x_{ij} ) \leq \frac{n\delta \sum_i \budgeti}{\xi}$;
    \item for any $i$, $p^\top \edwi = \budgeti$;
    \item $\sum_j p_j \geq 1$.
\end{enumerate}} 
Additionally, if ${u}^*_i - \delta < \tilde{u}_i \leq {u}^*_i$ for all $i$, then for any $\{x_i\}_i$ such that $\eqref{lp::cal-allocation}$ holds, $\{p^*_j\}_j$ is a solution to \eqref{dual-lp::utility-upper-used}, \eqref{lp::budget-constraint-used},  \eqref{lp::market-clear}, and \eqref{lp::endowment-budget}.
\end{theorem}

\medskip

For the overall algorithm, we approximate the utilities $u_i$ by $\bu_i^\xi$ as in \eqref{eq:bu}. We obtain the following similarly to Theorem~\ref{lem::fisher-algorithm-const-agent}. Note that the existence of an equilibrium is guaranteed for $\bu_i^\xi$ by Theorem~\ref{thm::exist-ad}. \hide{ even though an exact equilibrium may not exist for the $u_i$'s.}


\begin{theorem}
In an Arrow Debreu market with $n$ agents, $m$ items, 
and PLC utility functions $u_i$ with non-negative $\aijl$'s and $\sigma < 1$, we can compute a $\sigma$-approximate thrifty market equilibrium by solving $O(\left( \frac{n}{\sigma^2}\right)^n)$ linear programs, each in polynomial time in $n$, $m$, the bit-complexity of the  $\aijl$'s, $\log(1/\sigma)$ and $e_i$'s.
\end{theorem}

\subsection{Fixed number of items} 
In this section, we assume $\aijl \leq 1$ for any $i$, $j$ and $l$. The algorithm is more difficult as in Section~\ref{sec:Fisher-items}: we also rely on the robust approximate $\bu_i^\xi$ as in the fixed number of agents case.
The benefit of using $\bu_i^\xi$ is that we are able to lower bound the equilibrium prices:
\begin{lemma}\label{lem:min-price}
For $\xi>0$, let $(\{x^*_i\}_i, \{p^*_j\}_j)$ be a market equilibrium for the utilities $\bu_i^\xi$  with $\sum_j p^*_j=1$. Then
 $p^*_j \geq \frac{\xi}{m(m+\xi)}$ for all items $j$.
\end{lemma}
\begin{proof}
For a contradiction,
suppose there is an item $j$ such that $p^*_j < \frac{\xi}{m(m+\xi)}$. We must still have $p^*_j>0$: due to the form of $\bu_i^\xi$: the marginal utility of every good is strictly positive with respect to any bundle, and therefore all prices at equilibrium are strictly positive.

Let us select the most expensive item $j'$; by the assumption on the sum of the prices,  $p^*_{j'} \geq \frac{1}{m}$. Therefore, 
\begin{equation}\label{eq:p-p-p}
\frac{\xi}{m p^*_j} > \left({1 + \frac{\xi}{m}}\right)\cdot\frac{1}{p^*_{j'}}\end{equation}
 Consider an agent $i$ such that $x^*_{ij'} > 0$, and let $s_i = x^*_{ij'} p^*_{j'}>0$. Consider an alternative allocation $x'_i$ in which agent $i$ spends this $s_i$ amount of money on item $j$ instead of item $j'$: 
\begin{align*}
    x'_{ik} = \begin{cases}
    x^*_{ik} \, ,    & \quad \text{if } k \neq j, j'\, ,\\
    x^*_{ik} + \frac{s_i}{p^*_j}\, ,   & \quad \text{if } k = j\, ,\\
    0\, ,  & \quad \text{if } k = j'\, .\\
  \end{cases}.
\end{align*}
Then, for any $l$, from \eqref{eq:p-p-p} and the bound $\aijlp \le 1$ we see that
 \begin{align*}
    \left(\aijl + \frac{\xi}{m}\right) x'_{ij} + \bil &\geq \left(\aijl + \frac{\xi}{m}\right) x^*_{ij} + \bil +  \left(\aijl + \frac{\xi}{m}\right) \frac{s_i}{p^*_j} - \left(\aijlp + \frac{\xi}{m}\right) \frac{s_i}{p^*_{j'}} \\
    &\geq \left(\aijl + \frac{\xi}{m}\right) x^*_{ij} + \bil +   \frac{\xi}{m} \frac{s_i}{p^*_j} - \left(\amax + \frac{\xi}{m}\right) \frac{s_i}{p^*_{j'}} \\
    &> \left(\aijl + \frac{\xi}{m}\right) x^*_{ij} + \bil 
\end{align*}
This implies $x^*_i$ is not the optimal allocation for agent $i$, which gives a contradiction.
\end{proof}
Next, we give the algorithm.
\paragraph{Algorithm} 
 Similar to the Fisher-market, first, we enumerate ${p}_j = \delta k_j$ for all possible integers $k_j$ such that $0 \leq k_j \leq \left\lceil\frac{1}{\delta}\right\rceil + 1$ and $1 \leq \sum_j {p}_j \leq 1 + m\delta$. We then calculate the budget $\budgeti = {p}^\top \edwi$. With this budget and given the prices, we calculate the optimal utility $V^\xi_i(p,\budgeti)$ for $\bu^{\xi}_i(x_i)$; this can be found by solving a linear program.
 
Then, we calculate a feasible allocation $\{x_i\}_i$ by the following linear program:
\begin{equation}\label{lp::ad-const-item}
\begin{aligned}
    & \bu^{\xi}_i(x_i) \geq V^\xi_i( p,\budgeti) - \frac{\delta m  (m+ \xi)^2}{\xi}&\forall i\\
    & \sum_j x_{ij} {p}_j \leq \budgeti + \delta m &\forall i\\
    & \sum_i x_{ij} =1&\forall j \\
    &x_{ij} \geq 0 &\forall i,j. 
\end{aligned}
\end{equation}
If there is a feasible allocation $\{x_i\}_i$ to \eqref{lp::ad-const-item}, then we output the solution $(\{x_i\}_i, \{{p}_j\}_j)$.

The following lemma ensures that if $\{{p}_j\}_j$ is close to $\{p^*_j\}_j$, then $x^*$ is a feasible solution to \eqref{lp::ad-const-item}.
\begin{lemma}
If $p^*_j \leq {p}_j \leq p^*_j+ \delta$ for all $j$, then for any $i$, $V^\xi_i(p,\budgeti) \leq \bu^{\xi}_i(x^*_i)  + \frac{\delta m  (m + {\xi})^2}{\xi}$ and $\sum_j x^*_{ij} {p}_j \leq \budgeti + \delta m$. Additionally, $\sum_i x^*_{ij} = 1$.
\end{lemma}
\begin{proof}
Note that if $p^*_j \leq p_j \leq p^*_j +  \delta$, $\budgeti \leq (p^*)^\top e_i+ \delta m $. 
Since $\aijl\le 1$, the maximum amount of utility that can be obtained from $\varepsilon$ additional units of items is $\left(1+\frac{\xi}m\right)\varepsilon$. Moreover, $\delta m$ additional amount of money can buy at most $\delta m/\min_j p_j^*$ additional units of items.
Therefore,
\begin{align*}
    V^\xi_i(p,w_i) - \bu^{\xi}_i(x^*_i)&=V^\xi_i(p,w_i)-V^\xi_i(p,(p^*)^\top e_i ) \leq \left(1+\frac{\xi}m\right)\cdot \frac{\delta m}{\min_j p_j^*}\leq \frac{\delta m  (1 +\xi)^2}{\xi}\, ,
\end{align*}
where the last inequality follows by Lemma~\ref{lem:min-price}.

Also, $\sum_j  {p}_j  x^*_{ij}\leq \sum_j  p^*_j x^*_{ij} + \delta m = (p^*)^\top e_i + \delta m \leq  p^\top e_i+ \delta m = \budgeti + \delta_i$. Finally, $\sum_i x^*_{ij}=1$ is true by the market equilibrium definition, using that all equilibrium prices must be positive, as previously noted.
\end{proof}

Therefore, we obtain the following theorem. 
\begin{theorem}\label{thm::ad-fixed-items}
 Given an Arrow Debreu market with $n$ agents, $m$ items, 
and PLC utility functions $u_i$ such that $\aijl \in [0, 1]$ for any $i$, $j$ and $l$, we can compute a $(\sigma, \sigma)$-approximate thrifty market equilibrium by solving $O((12m^3 / \sigma^2)^m)$ linear programs, each in polynomial time in $n$, $m$, the bit-complexity of the $e_i$'s, $\aijl$'s, and $\log(1/\sigma)$. 
\end{theorem}
\begin{proof} Let $\xi = \frac{\sigma}{2}$, $\delta = \frac{\sigma^2}{12 m^3}$, and $(\{x_i\}_i, \{p_j\}_j)$ be the approximate equilibrium calculated by the algorithm. Then,
\begin{align*}
    u_i(x_i) &\geq u^{\xi}_i(x_i) - \xi ~~~~~\text{(by Lemma~\ref{lem::ad-u-1})}\\
&\geq  V^\xi_i( p,\budgeti) - \xi- \frac{\delta m  (m+ \xi)^2}{\xi} ~~~~~\text{(by \eqref{lp::ad-const-item})}\\
&\geq  V_i( p,\budgeti) - \xi- \frac{\delta m  (m+ \xi)^2}{\xi} ~~~~~\text{(by Lemma~\ref{lem::ad-u-1})}\\
&\geq  V_i( p,\budgeti) -\sigma\, .
\end{align*}
Additionally, $\sum_j x_{ij} p_j \leq \sum_j \edwij p_j + \delta m \leq \sum_j \edwij p_j + \sigma$. 
\end{proof}


\section{Existence of a market equilibrium}\label{sec::proof-existence-fisher}
In this section, we prove Theorem~\ref{thm::fisher-exists}.
We are done if a market equilibrium with $p\equiv 0$ exists. Otherwise, we define a correspondence $F(x,p): \R^{n\times m}\times\R^m\to  \R^{n\times m}\times\R^m$ where Kakutani's fixed point theorem can be used. 
The  domain $(x,p)$ for this correspondence is
\[S := \left\{ (x,p)\in \R^{n\times m}\times\R^m:\, x_{ij} \in [0, 1.1];\ \pj \geq 0;\  0 \leq \sum_j \pj \leq \sum_i \budgeti\right\}\enspace,\]

Let
\[
\Hd_i(p,w_i)=\arg\max\left\{u_i(x_i):\, x_i\in [0,1.1]^m\, , p^\top x_i\le w_i\right\}\, ,
\]
This differs from $D_i(p,w_i)$ used previously by restricting the bundles to $x_i\in [0,1.1]^m$. 

For allocations $x$ and prices $p$, we define $F(x,p)\subseteq  \R^{n\times m}\times\R^m$ as the set of pairs $(x', p')$ such that
\begin{itemize}
    \item $x'_i\in \Hd_i(p,w_i)$, and $x'_i$ minimizes $p^\top x'_i$ over $\Hd_i(p,w_i)$, i.e., $x'_i$ is a thrifty optimal bundle.
    \item $p' \in \arg \max \sum_{i,j}  p'_j x_{ij}\text{ s.t. } p'_j \geq 0 \text{ and }  0 \leq \sum_j p'_j \leq \max\{ \min_i \budgeti, \sum_{i,j}  p_j x_{ij}\} $. That is, $p'$ is supported only on the items in  maximum demand at $x$. 
\end{itemize}
\begin{lemma}
If there is no market equilibrium with  $p\equiv 0$, then any fixed point of $F(x, p)$ is a market equilibrium.
\end{lemma}
\begin{proof}
Let $(\{x_i\}_{i}, \{\pj\}_j)$ be a fixed point, i.e., $(x,p) \in F(x,p)$. By the budget constraint, $\sum_{i,j}  p_jx_{ij} \leq \sum_i \budgeti$. Note that by the second condition in the definition of $F(x,p)$, we must have either $\sum_j p_j=\min_i \budgeti$ or $\sum_j p_j =\sum_{i,j}  p_j x_{ij}$.

\paragraph{Case I: $\min_i \budgeti \leq \sum_{i,j} p_j  x_{ij}.$}
In this case, $\sum_j p_j = \sum_{i,j}  p_j x_{ij}$, and
 there exist at least one $x_{ij} > 0$. Therefore,
 $\sum_j p_j = \sum_{i,j}  p_j x_{ij} = \left(\max_j \sum_i x_{ij}\right) \sum_j p_j$,
 implying $\max_j \sum_i x_{ij} = 1$ and  $p_j > 0$ only if $\sum_i x_{ij} = 1$, which is the market clearing condition. Let us show that every agent is getting an optimal bundle, at the minimum possible price: $x_i\in D_i(p,w_i)$ and $p^\top x_i=C_i(p,w_i)$.
\begin{itemize}
    \item If $x_i\notin D_i(p,w_i)$, then there exists $\tilde{x}_i$ within budget constraints such that $u_i(\tilde{x}_i) > u_i(x_i)$. Since $x_i\in [0,1]^m$, 
    we can find an $\varepsilon>0$ such that $(1 - \varepsilon) x_i + \varepsilon \tilde{x}_i \leq 1.1$. By convexity, $u_i((1 - \varepsilon) x_i  + \varepsilon \tilde{x}_i) > u_i(x_i)$, contradicting $x_i \in \Hd(p,w_i)$.
    \item If $x_i\in D_i(p,w_i)$ but $p^\top x_i>C_i(p,w_i)$, then there exists $\tilde{x}_i$ such that $p^\top \tilde{x}_i < p^\top x_i$ and $u_i(\tilde{x}_i) = u_i(x_i)$. Similarly, for a small positive $\epsilon$ such that $(1 - \epsilon) x_i + \epsilon \tilde{x}_i \leq 1.1$,  $u_i((1 - \epsilon) x_i + \epsilon \tilde{x}_i) \geq u_i(x_i)$ and $p^\top ((1 - \epsilon) x_i + \epsilon \tilde{x}_i) < p^\top x_i$, which violates the fact that $p^\top x_i$ minimized over $\Hd(p,w_i)$.
\end{itemize}
Therefore, in this case, $(\{x_i\}_{i}, \{\pj\}_j)$ gives a thrifty market equilibrium.

\paragraph{Case II: $\min_i \budgeti > \sum_{i,j} x_{ij} p_j$.} In this case,  $\sum_j p_j=\min_i \budgeti$.
We show that  $p\equiv 0$ is a market equilibrium, contrary to the assumption. 
Indeed, either $\sum_i x_{ij} = 0$ for all $j$, or 
\[
\sum_j p_j = \min_i \budgeti > \sum_{i, j} x_{ij}p_j =\left( \max_j \sum_i x_{ij} \right) \sum p_j\, .
\]
 In both cases, $\sum_i x_{ij} < 1$ for all $j$ and no one spent out their budget, $\sum_j x_{ij} p_j < \min_i \budgeti$. This implies all agents get their favorite bundle (without the budget constraint) while the total demand of each good is less than $1$.
 However, in such a case $p\equiv 0$ gives an equilibrium.
\end{proof}
\begin{lemma}
There exists a fixed point of correspondence $F(x, p)$.
\end{lemma}
\begin{proof}
We show this using the Kakutani's fixed point theorem~\cite{kakutani1941generalization}. 

First, it is easy to check for any $(x, p)$, $F(x,p)\neq\emptyset$ and convex. For $x$, these follow since by the regularity assumption $u_i(0)=0$; and therefore $H_i(p,w_i)$ amounts to maximizing a concave function over a nonempty compact 
set. The nonemptyness and convexity of the feasible $p'$ values is immediate.

Finally, we need to show that the map $F$ is upper hemicontinuous.  Consider a sequence of allocations and prices such that $\lim_{r \rightarrow \infty} (x^{(r)}, p^{(r)}) = (x, p)$, $({x'}_i^{(r)}, p'^{(r)}) \in F(x^{(r)}, p^{(r)})$, and $\lim_{r \rightarrow \infty}  {x'}_i^{(r)} = x'_i$,
$\lim_{r \rightarrow \infty}  {p'}_i^{(r)} = p'_i$. 
Upper hemicontinuity requires $(x'_i,p'_i) \in F(x, p)$. 

For the prices $p'$, this easily follows by the Maximum theorem\footnote{$\sum_{ij} x_{ij} p_j$ is a continuous function of $x$ and $p$; and $U(x, p) = \{p'_j \geq 0;  0 \leq \sum_j p'_j \leq \max\{ \min_i \budgeti, \sum_{i,j} x_{ij} p_j \}$ is an upper and lower continuous correspondence.}. So, we only need to show that $x'$ is also upper hemicontinuous. 

We prove this by a contradiction.
 Let $(\tilde{x}_i,p'_i)\in F(x, p)$, There are two cases.
\begin{itemize}
    \item $u_i(\tilde{x}_i) > u_i(x'_i)$. We claim that $u_i({x'_i}^{(r)})$ will be no smaller than $\frac{u_i(\tilde{x}_i) + u_i(x'_i)}{2}$ if $r$ is big enough, which provides a contradiction.\footnote{Since ${x'_i}^{(r)} \in F(x^{(r)}, p^{(r)})$, ${x'_i}^{(r)} \in K_i$, the domain of $u_i$. By regularity, $K_i$ is closed, and $u_i$ is continuous of $K_i$, therefore  $\lim_{r \rightarrow \infty} u_i({x'_i}^{(r)}) = u_i(x'_i)$.} This is because we can find a small $\varepsilon$, such that for $y_i=(1 - \varepsilon)\tilde x_i$, we have $u_i(y_i)\ge  u_i(\tilde{x}_i) + \varepsilon u_i(0) \geq \frac{u_i(\tilde{x}_i) + u_i(x'_i)}{2}$. Additionally, we know that $p^\top y_i =(1 - \varepsilon) p^\top \tilde{x}_i$, which implies that $y^{(i)}$ is a feasible solution for iterate $p^{(r)}$ if $p^{(r)}$ is close enough to $p$. Therefore, the $u_i({x'_i}^{(r)}) \geq u_i(y_i) \geq \frac{u_i(\tilde{x}_i) + u_i(x'_i)}{2}$.
    \item $u_i(\tilde{x}_i) = u_i(x'_i)$ but $p^\top \tilde{x}_i < p^\top x'_i$. In this case, for those $p^{(r)}$ which is close enough to $p$, $\tilde{x}_i$ is also a budget feasible allocation which maximized the utility function. This is because $p^\top \tilde{x}_i < p^\top x'_i  \leq \budgeti$ and any better solution of $p^{(r)}$ will provide a solution (a convex combination of this better solution and $\tilde{x}_i$) of strictly better utility than $\tilde{x}_i$ at price $p$. Since $\tilde{x}_i$ is a budget feasible allocation which maximized the utility function for those $p^{(r)}$, $\sum_j {x'}^{(r)}_{ij} p^{(r)}_j \leq \sum_j \tilde{x}_{ij} p^{(r)}_j$. However, $ \sum_j x'_{ij} p_j = \lim_{r \rightarrow \infty} \sum_j {x'_{ij}}^{(r)} p^{(r)}_j \leq \lim_{r \rightarrow \infty} \sum_j \tilde{x}_{ij} p^{(r)}_j = \sum_j \tilde{x}_{ij} p_j$ which contradicts the fact that $p^\top \tilde{x}_i < p^\top x'_i$.
\end{itemize}
This completes the proof of Theorem~\ref{thm::fisher-exists}.
\end{proof}

\bibliographystyle{abbrv}
\bibliography{references}
\end{document}